\documentclass{article}

\usepackage{longtable}
\usepackage[english]{babel} 
\usepackage{amsmath,amsfonts,amsthm,bm} 
\usepackage{graphicx}
\usepackage{gensymb}
\usepackage{dcolumn}
\usepackage{bm}
\usepackage{color}
\usepackage[autostyle]{csquotes}
\usepackage{tikz}
\usepackage{setspace}
\usepackage{multirow}
\usepackage{authblk}
\DeclareMathOperator{\sign}{sign}
\usepackage[latin1]{inputenc}
\usepackage{hyperref}

\usepackage{geometry}
\begin{document}

\title{Infinitely Many Multipulse Solitons of Different Symmetry Types \\ in the Nonlinear Schr\"{o}dinger Equation  with Quartic Dispersion }

\author[1,3]{Ravindra Bandara\thanks{ravindra.bandara@auckland.ac.nz}}
\author[1,3]{Andrus Giraldo}
\author[2,3]{Neil G. R. Broderick}
\author[1,3]{Bernd Krauskopf}
\affil[1]{Department of Mathematics, University of Auckland, Auckland 1010, New Zealand.}
\affil[2]{Department of Physics, University of Auckland, Auckland 1010, New Zealand.}
\affil[3]{The Dodd-Walls Centre for Photonic and Quantum Technologies, New Zealand.}

\date{\today}
\maketitle

\begin{abstract}
We show that the generalised nonlinear Schr\"{o}dinger equation (GNLSE) with quartic dispersion supports infinitely many multipulse solitons for a wide parameter range of the dispersion terms. These solitons exist through the balance between the quartic and quadratic dispersions with the Kerr nonlinearity, and they come in infinite families with different signatures. A travelling wave ansatz, where the optical pulse does not undergo a change in shape while propagating, allows us to transform the GNLSE into a fourth-order nonlinear Hamiltonian ordinary differential equation with two reversibilities. Studying families of connecting orbits with different symmetry properties of this reduced system, connecting equilibria to themselves or to periodic solutions, provides the key to understanding the overall structure of solitons of the GNLSE. Integrating a perturbation of them as solutions of the GNLSE suggests that some of these solitons may be observable experimentally in photonic crystal wave-guides over several dispersion lengths.
\end{abstract}


\section{INTRODUCTION}

Recently, Blanco-Redondo \textit{et al.}~experimentally discovered pure quartic solitons (PQS) in a silicon photonic crystal waveguide \cite{blanco2016pure}. 
Such PQS exist due to a balance between negative quartic dispersion and the Kerr nonlinearity --- unlike conventional optical solitons which balance quadratic dispersion and nonlinearity. This balance through the quartic dispersion allows for an unusual scaling of the pulse width with the power, which makes them attractive for short-pulse lasers \cite{blanco2016pure}. Furthermore, these new solitons have decaying oscillating tails. Pure quartic solitons have been the focus of several recent studies, both experimental and theoretical \cite{tam2018solitary, tam2019stationary, GDK}.

It is well known that the generalised nonlinear Schr\"{o}dinger equation (GNLSE) can be used to model optical pulse propagation in fibres in a variety of different regimes, including the case of higher-order dispersion. An analytic solution in the presence of both quartic and quadratic dispersion was found by Karlson and H\"{o}\"{o}k \cite{karlsson1994soliton} in the form of a squared hyperbolic secant pulse shape. It exists in the situation when both the quadratic and the quartic dispersion coefficients, $\beta_2$ and $\beta_4$, are negative. Furthermore, the Karlson and H\"{o}\"{o}k solution family does not possess oscillating decaying tails; thus, it does not describe PQS. Subsequently, Pich\'{e} \textit{et al}.~\cite{Piche:s} numerically studied the effect of third-order dispersion together with second and fourth-order dispersion. With numerical simulations, they found that, when a weak third-order dispersion is introduced, the temporal profile and the peak power of the soliton remain unchanged. Akhmediev \textit{et al}.~\cite{AKHMEDIEV1994540} found that, when both $\beta_{2}$ and $\beta_{4}$ are negative, solitons have either exponentially decaying tails or oscillating tails, depending on the soliton propagation constant (nonlinear shift of the wave number). Interaction of solitons with oscillating tails was numerically studied by Akhmediev and Buryak \cite{AKH}. They found that the oscillating tails of a soliton establish a potential barrier between neighbouring solitons during their interactions, preventing two adjacent solitons from combining. Furthermore, they investigated the bound states of two or more solitons when the single soliton has oscillating tails \cite{PhysRevE.51.3572}. Roy and Biancalana \cite{PhysRevA.87.025801} demonstrated that, when both $\beta_{2}$ and $\beta_{4}$ are negative, it is possible to observe solitons in silicon-based slot waveguides. Although there have been many studies of solitons in the presence of both quartic and quadratic dispersions, all these works considered the situation when both $\beta_{2}$ and $\beta_{4}$ are negative. More recently Tam \textit{et al}.~\cite{tam2018solitary,GDK} considered the case when $\beta_{4}$ is negative while $\beta_{2}$ can have either sign. They numerically found that single-hump solitons exist for some positive $\beta_{2}$ values as well. Furthermore, they explained that the decay rate of the solitons decreases as $\beta_{2}$ increases.

In this paper, we perform a detailed analysis of the GNLSE to find solitons of different types (multi-hump solitons) with different symmetry properties, beyond the one-hump soliton obtained in \cite{GDK} and for different signs of the quadratic dispersion term. By taking a dynamical system approach, we show that the GNLSE supports infinitely many multi-hump solitons in the presence of both quartic and quadratic dispersion. We consider the situation when $\beta_{4}$ is negative, and show that these multi-hump solitons exist not only when $\beta_{2}$ is negative. More specifically, we extend the work in \cite{PhysRevE.51.3572} and \cite{GDK} and present the parameter range for which these multi-hump solitons also exist when $\beta_2$ is positive. Moreover, we show that, apart from symmetric and antisymmetric solitons, there also exist nonsymmetric, symmetry-broken solitons, which are distinct from a union of fundamental solitons. For fixed negative $\beta_{4}$, we determine the parameter intervals of $\beta_{2}$ over which the different types of solitons exists. This is achieved via a travelling wave ansatz that transforms the GNLSE into a fourth-order Hamiltonian ordinary differential equation (ODE) with two reversible symmetries. Solitons of the GNLSE are then identified as homoclinic solutions to the origin of this ODE, which we find and track with state-of-the art continuation techniques. We discuss a connection of our findings with results obtained, for example, for the Swift-Hohenberg equation \cite{articleBruke} and the Lugiato-Lefeverer equation \cite{articleParra}, and demonstrate that the overall bifurcation structure of the GNLSE can be characterised as truncated homoclinic snaking. Moreover, we show that connecting orbits from the origin to periodic orbits (also referred to as EtoP connections) with different symmetry properties organise different infinite families of multi-hump solitons. The structure of periodic orbits is discussed briefly to show that infinitely many of them create different connections to the origin and, hence, associated families of solitons. Importantly, our results apply to any negative values of the quartic dispersion term via a suitable transformation. Finally, we investigate the evolution of these solitons along a waveguide via the integration of the GNLSE with a split-step Fourier method; here we consider symmetric and non-symmetric solitons for $\beta_2 = 0$ and also for different signs of $\beta_2$. Our numerical simulations indicate that some of the multi-hump solitons are only weakly unstable and may propagate effectively unchanged over many dispersion lengths; hence, they might be observable in careful experiment with currently available waveguides.

The paper is structured as follows. In Sec.~\ref{sec:analysis}, we introduce the GNLSE, and show how it can be transformed into an ODE; we then discuss the special mathematical properties of this ODE, present its local bifurcation analysis and also discuss the role of its homoclinic solutions as solitons of the GNLSE. In Sec.~\ref{sec:bvp}, we set up suitable boundary value problems to find and then continue homoclinic solutions, periodic solutions and EtoP connections. In Sec.~\ref{sec:connorbits}, we present one-by-one families of homoclinic solutions of different symmetry types, show over which $\beta_2$-range they exist and discuss the connection with homoclinic snaking. Section~\ref{sec:parplane} then shows that the respective homoclinic solutions occur along parabolas in different parameter planes. In Sec.~\ref{sec:perorbits} we show that there are infinitely many periodic orbits that create families of connecting orbits and, hence, a menagerie of solitons with different signatures. In Sec.~\ref{sec:simulations}, we present some simulations of the GNLSE that demonstrate that, while only the single-pulse primary soliton is stable, certain multi-hump solitons are only weakly unstable and, thus, may be observable in an experimental settings. Finally, a discussion of the results and an outlook to future research are presented in Sec.~\ref{sec:conclusions}.

\section{MATHEMATICAL ANALYSIS}
\label{sec:analysis}

Pulse propagation along an optical fibre under the influence of quadratic dispersion, quartic dispersion and the Kerr nonlinearity is governed by the
GNLSE \cite{agrawal2000nonlinear}
       \begin{equation}
       \frac{\partial A}{\partial z}=i \gamma |A|^{2} A - i \frac{\beta_{2}}{2}\frac{\partial ^{2} A}{\partial t^{2}}+i\frac{\beta_{4}}{24}\frac{\partial ^{4} A}{\partial t^{4}}.
       \label{gnse}
       \end{equation}
Here, $A(z,t)$ is the slowly varying complex pulse envelope, $z$ is the propagation distance, $t$ is the time in a co-moving frame of the pulse, $\gamma$ is the coefficient of the nonlinearity, and $\beta_{2}$ and $\beta_{4}$ are the quadratic and quartic dispersion coefficients, respectively. In order to focus on the essential interactions between dispersion and nonlinearity that drive soliton formation, Eq.~\eqref{gnse} does not include losses and higher-order terms such as the Raman effect. 

\begin{table*}
\begin{center}
\begin{tabular}{ccccc}
  Quadrant   & Transformation & Non-dimensionalised GLNSE \\ \hline  \\
(a) $\beta_{2}>0$, $\beta_{4}>0$ &       & $\cfrac{\partial U}{\partial x}=i |U|^{2} U -i \cfrac{\partial ^{2} U}{\partial \tau^{2}}+i\cfrac{\partial ^{4} U}{\partial \tau^{4}}$   \\ \\
(b) $\beta_{2}<0$, $\beta_{4}>0$  & $U=\sqrt{\cfrac{\sign(\beta_{4})\beta_{4}\gamma}{6\beta_{2}^{2}}}A$, $x=\cfrac{6\beta_{2}^{2}}{\sign(\beta_{4})\beta_{4}}z$   & $\cfrac{\partial U}{\partial x}=i |U|^{2} U +i \cfrac{\partial ^{2} U}{\partial \tau^{2}}+i\cfrac{\partial ^{4} U}{\partial \tau^{4}}$   \\ \\
(c) $\beta_{2}>0$, $\beta_{4}<0$  & $\tau=\sqrt{\cfrac{12\sign(\beta_{2})\beta_{2}}{\sign(\beta_{4})\beta_{4}}} t$  & $\cfrac{\partial U}{\partial x}=i |U|^{2} U - i \cfrac{\partial ^{2} U}{\partial \tau^{2}}-i\cfrac{\partial ^{4} U}{\partial \tau^{4}}$ \\ \\
(d) $\beta_{2}<0$, $\beta_{4}<0$  &       & $\cfrac{\partial U}{\partial x}=i |U|^{2} U + i \cfrac{\partial ^{2} U}{\partial \tau^{2}}-i\cfrac{\partial ^{4} U}{\partial \tau^{4}}$ \\ \\
\end{tabular}
\caption{\label{tab:reductions} Different transformations to reduce the GNLSE to a parameter-free form in each quadrant of the $(\beta_{2}, \beta_{4})$-plane.}
\end{center}
\end{table*}

We note that it is possible to reduce Eq.~\eqref{gnse} to a parameter-free form by using a rescaling transformation. However, as shown in Table~\ref{tab:reductions}, depending on the sign of $\beta_{2}$ and $\beta_{4}$, it is necessary to consider different transformations. That is, each quadrant of the $(\beta_{2},\beta_{4})$-plane has a corresponding transformation that reduces Eq.~\eqref{gnse} into a parameter-free form. In particular, Akhmediev \textit{et al}.~\cite{AKHMEDIEV1994540} considered the transformation of case (d) in Table~\ref{tab:reductions}; thus, they considered the situation when both $\beta_{2}$ and $\beta_{4}$ are negative. Furthermore, all the transformations are undefined when $\beta_{2}=0$ or $\beta_4=0$, and there is no continuous transition between different signs of $\beta_2$ and $\beta_4$. Therefore, none of the reduced PDEs are able to describe the case of PQS. For the purpose of this paper, we are interested in the transition between $\beta_{2}<0$, $\beta_{2}=0$ (PQS) and $\beta_{2}>0$. Thus, we consider the original GNLSE without reducing the parameters first. We stress that our results are general because they can be mapped to any specific parameter region by the corresponding transformations for given signs of $\beta_2$ and $\beta_4$.

When solving the GNLSE one looks for solutions, where the pulse is stationary and does not change with propagation distance. We study here such travelling wave solutions of the form
    \begin{equation}
        A(z,t)=u(t)e^{i\mu z},
        \label{ansatz}
        \end{equation}
 where $u(t)$ is the temporal profile of the pulse and $\mu$ is the soliton propagation constant (nonlinear shift of the wavenumber) \cite{AKHMEDIEV1994540}. Note that the intensity of such solutions, $|A(t)|^2 =u(t)^{2}$, is unchanged during propagation. After substituting Eq.~\eqref{ansatz} into Eq.~\eqref{gnse} we obtain the fourth-order nonlinear ODE
        \begin{equation}
        \frac{\beta_{4}}{24} \frac{d^{4}u}{dt^{4}}-\frac{\beta_{2}}{2} \frac{d^{2}u}{dt^{2}}-\mu u+\gamma u^3=0.
        \label{ode}
        \end{equation}
By introducing the new variables $u_{1}, u_{2}, u_{3}$ and $u_{4}$ such that $\mathbf{u}=(u_1,u_2,u_3,u_4)=\left(u,\frac{du}{dt},\frac{d^{2}u}{dt^{2}},\frac{d^{3}u}{dt^{3}}\right)$, Eq.~\eqref{ode} can be written as the system of four first-order ODEs
        \begin{equation}
	\cfrac{d\mathbf{u}}{dt}=f({\bf{u}},\zeta)=\left( {\begin{array}{cc}u_{2} \\u_{3}\\u_{4}\\ \cfrac{24}{\beta_{4}}\left(\cfrac{\beta_{2}}{2}u_{3}+\mu u_{1}-\gamma u_{1}^3  \right) \\\end{array} } \right),
	\label{system}
	\end{equation}\\
where  $\zeta=(\beta_{2},\beta_{4},\gamma,\mu)\in \mathbb{R}^{4}$. Note that system \eqref{system} is reversible \cite{champneys1993hunting} under the transformations
	\begin{itemize}
	\item[] $R_{1}: (u_{1},u_{2},u_{3},u_{4}) \rightarrow (u_{1},-u_{2},u_{3},-u_{4})$ and 
	\item[] $R_{2}: (u_{1},u_{2},u_{3},u_{4}) \rightarrow (-u_{1},u_{2},-u_{3},u_{4}).$
	\end{itemize}
 This means that if $\mathbf{u}(t)$ is a solution then both $R_{1}(\mathbf{u}(-t))$ and $R_{2}(\mathbf{u}(-t))$ are also solutions of system \eqref{system}. Throughout the paper, we refer to $R_1(\mathbf{u}(-t))$ and $R_2(\mathbf{u}(-t))$ as the $R_1$- and $R_2$-counterpart of $\mathbf{u}(t)$, respectively.
Furthermore, $S=R_{1} \circ  R_{2}=R_{2} \circ R_{1}$, is the state-space symmetry 
	\begin{itemize}
	\item[] $S: (u_{1},u_{2},u_{3},u_{4}) \rightarrow (-u_{1},-u_{2},-u_{3},-u_{4})$
	\end{itemize}
of system \eqref{system}, which is point reflection in the origin $\mathbf{0}$.
The set of points that are left invariant under $R_{1}$ or $R_{2}$ are known as symmetric or reversibility sections of system \eqref{system}; they are
	\begin{itemize}
	\item[] $\Sigma_{1}=\{\mathbf{u}\in \mathbb{R}^{4} : u_{2}=u_{4}=0  \}$ and
	\item[] $\Sigma_{2}=\{\mathbf{u}\in \mathbb{R}^{4} : u_{1}=u_{3}=0  \}$,
	\end{itemize}
respectively.
Note that the origin $\mathbf{0}$ is the only point that belongs to both $\Sigma_{1}$ and $\Sigma_{2}$, that is, it is the only point that is invariant under $S$. A solution trajectory $\mathbf{u}(t)$ of system~(4) is called \emph{symmetric} if it satisfies either $R_{1}(\mathbf{u}(-t))=\mathbf{u}(t)$ or $R_{2}(\mathbf{u}(-t))=\mathbf{u}(t)$. One can show that if $\mathbf{u}(t)$ is a symmetric solution, then there exists a time  $t^{*}\in\mathbb{R}$ such that $\mathbf{u}(t^{*}) \in \Sigma_{1}$ or $\mathbf{u}(t^{*}) \in \Sigma_{2}$, that is, $\mathbf{u}(t)$ intersects $\Sigma_1$ or $\Sigma_2$.  To distinguish  the invariance between the two reversibility conditions $R_1$ and $R_2$, we refer to a solution that is only invariant under $R_1$ as a $R_1$-symmetric solutions of system \eqref{system}, and similarly define $R_2$-symmetric solutions. Furthermore, if a solution is invariant under both $R_1$ and $R_2$, then we refer to it as $R^*$-symmetric. Notice that if a solution is $R^{*}$-symmetric, then it is invariant under $S$; however, invariance under $S$ does not necessarily imply $R^*$-symmetry. Lastly, a solution that is neither invariant under $R_1$ nor $R_2$ is referred to as non-symmetric. 

System \eqref{system} can be transformed into the form of a Hamiltonian system by the change of coordinates  
$$\mathbf{p}=(p_{1},p_{2})=(u_{2},u_{4}),$$
$$\mathbf{q}=(q_{1},q_{2})=\left(\frac{-12\beta_{2}}{\beta_{4}}u_{1}+u_{3},u_{1}\right),$$
where $\mathbf{p}$ and $\mathbf{q}$ are the generalised position and momentum coordinates. With this transformation, system \eqref{system} can be written as
	\begin{equation}
	\left( {\begin{array}{cc} \vspace{0.15cm} \cfrac{dq_{1}}{dt} \\ \vspace{0.1cm} \cfrac{dq_{2}}{dt} \\ \vspace{0.15cm}  \cfrac{dp_{1}}{dt}\\ \vspace{0.1cm}  \cfrac{dp_{2}}{dt}  \\\end{array} } \right)=\left( {\begin{array}{cc} \vspace{0.15cm} \cfrac{-12\beta_{2}}{\beta_{4}}p_{1}+p_{2}  \\ \vspace{0.15cm} p_{1} \\ \vspace{0.15cm} q_{1}+\cfrac{12\beta_{2}}{\beta_{4}}q_{2}\\ \vspace{0.15cm}  \cfrac{24}{\beta_{4}}\left(\cfrac{\beta_{2}}{2}\left(q_{1}+\cfrac{12\beta_{2}}{\beta_{4}}q_{2}\right)+\mu q_{2}-\gamma q_{2}^{3}\right) \\\end{array} } \right),
	\label{systemHam}
	\end{equation}\\
which satisfies the well-known Hamiltonian equations \cite{champneys1993hunting}
	\begin{equation}
	\frac{d\mathbf{q}}{dt}=\frac{\partial \widehat{H}}{\partial \mathbf{p}}, \quad \frac{d\mathbf{p}}{dt}=-\frac{\partial \widehat{H}}{\partial \mathbf{q}}.
	\label{Ham}
	\end{equation}
Here the conserved quantity or energy is
	\begin{eqnarray}
	\widehat{H}(\mathbf{p},\mathbf{q})&=&p_{1}p_{2}-\frac{6\beta_{2}}{\beta_{4}}p_{1}^{2}
	+\frac{1}{2}\left(q_{1}
	+\frac{12\beta_{2}}{\beta_{4}}q_{2}\right)^{2}\nonumber\\
	&&+
	\left(\frac{6\gamma q_{2}^{4}-12\mu q_{2}^{2}}{\beta_{4}}\right),
	\end{eqnarray}
as obtained from system \eqref{systemHam} by integration with respect to $\mathbf{p}$ and $\mathbf{q}$.
This expression can be written in original coordinates as 
	 \begin{equation}
        H(\mathbf{u})=u_{2}u_{4}-\frac{1}{2}u_{3}^{2}-\left(\frac{6\beta_{2}u_{2}^{2}-6\gamma u_{1}^{4}+12\mu u_{1}^{2}}{\beta_{4}}\right),
        \label{Hamiltonian}
        \end{equation}
and it is a conserved quantity along solution trajectories of system~\eqref{system}. We remark that one can also derive Eq.~\eqref{Hamiltonian} by using the general expression for fourth-order reversible systems provided in \cite{champneys1998homoclinic}. Notice that, if a solution trajectory  converges backward or forward in time to an equilibrium $u_0$ of system~\eqref{system} with energy $H(\mathbf{u_0})$, then the solution trajectory has energy $H(\mathbf{u_0})$ for all times. 

We now focus our attention on the equilibria of system \eqref{system}. The origin $\mathbf{0}=(0,0,0,0)$ is an equilibrium for any parameter value. It undergoes a pitchfork bifurcation at $\mu=0$, which creates two equilibria $\mathbf{E_{\pm}}=\left(\pm \sqrt \frac{\mu}{\gamma},0,0,0\right)$ for $\mu \gamma>0$. These are the only equilibria of system \eqref{system}. Note that $\mathbf{0} \in \Sigma_{1} \cap \Sigma_{2}$, while $\mathbf{E_{\pm}}$ lie only in $\Sigma_1$. Hence, all equilibria are symmetric: $\mathbf{0}$ is invariant under both $R_1$ and $R_2$, and $\mathbf{E_{\pm}}$ are invariant under $R_1$ only. Note from Eq.~\eqref{Hamiltonian} that, for any choice of the parameters, $\mathbf{0}$ always lies in the zero energy level. This is not the case for the other two equilibria $\mathbf{E_{\pm}}$ for which $H(\mathbf{E_{\pm}})=-\frac{6\mu^2}{\gamma\beta_{4}}$. Hence, $H(\mathbf{E_{\pm}}) \neq 0$ whenever $\mu \neq 0$ (and $ \beta_{4},\gamma \neq 0$), so that the equilibria $\mathbf{0}$ and $\mathbf{E_{\pm}}$ do not lie in the same energy level. Thus, there cannot be a connecting trajectory between them. 

We now focus our attention on \emph{homoclinic solutions}, which are trajectories of system \eqref{system} that converge to the same equilibrium both forward and backward in time. Homoclinic solutions are sought because they correspond to solitons of the GNLSE. Since, the solitons of the GNLSE converge to $0$, in both forward and backward in time, we only consider homoclinic solutions to $\mathbf{0}$ as it is the only equilibrium with $u=u_1=0$. 

Homoclinic solutions in fourth-order, reversible and Hamiltonian systems have been studied, for example, in \cite{devaney1977blue, devaney1976, champneys1998homoclinic, amick_toland_1992,champneys1993hunting,champneys1993bifurcation,HOMBURG2010379}. In particular, results on four-dimensional reversible systems have been developed and applied in the analysis of a system that describes the dynamics of an elastic strut \cite{champneys1998homoclinic, amick_toland_1992,champneys1993bifurcation, champneys1993hunting}. It is the case that symmetric homoclinic solutions in fourth-order reversible systems persist when a suitable parameter is changed \cite{devaney1977blue, devaney1976, champneys1998homoclinic}. This is true for both reversible and non-reversible Hamiltonian systems. It has also been proved that each symmetric homoclinic solution in a reversible system is accompanied, for fixed parameter values, by a one-parameter family of periodic solutions with minimal period $T_0$. As the period growths to infinity, that is $T_0 \rightarrow \infty$, periodic solutions accumulate on a symmetric homoclinic solution. The periodic solutions of these families lie in different Hamiltonian energy levels, and some of them lie in the energy level with $H=0$. As opposed to the symmetric case, non-symmetric homoclinic solutions in reversible systems do not persist as a suitable parameter is changed; however, they persist in systems that are both reversible and Hamiltonian \cite{champneys1998homoclinic,ELVIN2010537} as is the case for system \eqref{system}. It is a special property of systems that are both reversible and Hamiltonian that homoclinic solutions persist as codimension-zero phenomena. It has been proven in \cite{champneys1993bifurcation} that a transition of the equilibrium from a real saddle (spectrum with only real eigenvalues) to a saddle-focus (spectrum with only complex conjugate eigenvalues), as a parameter is changed, is an organising centre for the creation of infinitely many symmetric homoclinic solutions, provided the following conditions are satisfied:
\begin{itemize}
\item[(1)] the ODE is fourth-order, reversible and Hamiltonian,
\item[(2)] there exists a symmetric homoclinic solution at the moment of the transition of the equilibrium.
\end{itemize}
One refers to this transition as the Belyakov-Devaney (BD) bifurcation \cite{devaney1977blue,articleParra,champneys1998homoclinic,haragus2010local}. To see whether the second condition is satisfied for system \eqref{system}, we first focus our attention on the eigenvalues of its equilibria in different parameter regimes. The eigenvalues of the linearisation around a symmetric equilibrium of a reversible system have generically one of the following forms  \cite{champneys1993hunting}:
\begin{itemize}
\item[(I)]{two eigenvalues are $\pm \lambda_1$ and the other two are $ \pm \lambda_2$, where  $\lambda_1, \lambda_2 \in \mathbb{R}$};
\item[(II)]{two eigenvalues are $\pm \lambda$ and the other two are $\pm  \lambda^{*}$, where  $\lambda \in \mathbb{C}$ with $\operatorname{Re}(\lambda) \neq 0$ and $\operatorname{Im}(\lambda) \neq 0$};
\item[(III)]{two eigenvalues are $\pm  i \lambda_1$ and the other two are $\pm i  \lambda_2$, where  $\lambda_1, \lambda_2 \in \mathbb{R}$};
\item[(IV)]{two eigenvalues are $\pm  \lambda_1$ and  the other two are $\pm i  \lambda_2$, where  $\lambda_1, \lambda_2 \in \mathbb{R}$}.
\end{itemize}
For system \eqref{system}, one can obtain an analytical expression for the eigenvalues of the linearisation around $\mathbf{0}$ and $E_{\pm}$. The eigenvalues of $\mathbf{0}$ are given by \cite{GDK, champneys1998homoclinic}:
	\begin{equation}	\lambda_{\mathbf{0}}^{2}=\frac{6\beta_{2}}{\beta_{4}}\left(1\pm\sqrt{1+\frac{2\beta_{4}}{3\beta_{2}^{2}}\mu}\right),
	\label{EV}
	\end{equation}
while the eigenvalues of $E_{\pm}$ are given by
	\begin{equation}	\lambda_{E_{\pm}}^{2}=\frac{6\beta_{2}}{\beta_{4}}\left(1\pm\sqrt{1-\frac{4\beta_{4}}{3\beta_{2}^{2}}\mu}\right).
	\label{EV E}
	\end{equation}
	
\begin{figure}[t!]
  	 \centering
    	 \includegraphics[scale=1.3]{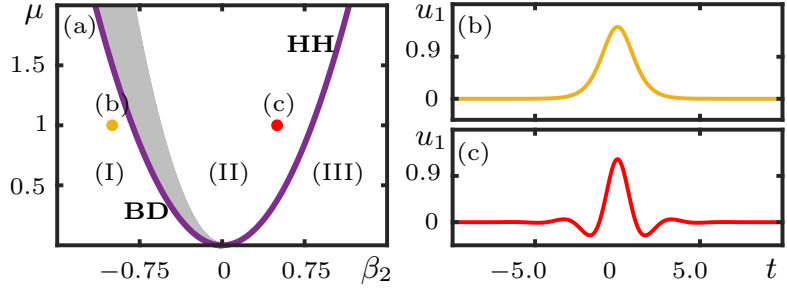}
   	 \caption{Panel (a) shows the bifurcation diagram in the $(\beta_{2}, \mu)$- parameter plane for $\beta_{4}=-1$ and $\gamma=1$. The purple parabola denotes the boundary between the equilibrium $\mathbf{0}$ having real and complex eigenvalues, given by BD and HH bifurcation; in the grey shaded region close to the purple parabola, infinitely many symmetric homoclinic solutions are expected to exist.  Panel (b) and (c) show the temporal traces of the primary homoclinic solution of system~\eqref{system} for $(\beta_{2}, \beta_{4}, \gamma, \mu)=(-1, -1, 1, 1)$ and $(\beta_{2}, \beta_{4}, \gamma, \mu)=(0.4, -1, 1, 1)$, at the yellow and orange dots in panel~(a), respectively.} 
\label{fig:basic}
\end{figure}

Since we are interested in homoclinic solutions to $\mathbf{0}$, we focus our attention on Eq.~\eqref{EV}. Notice that the expression inside the square root defines a parabola in the $(\beta_2,\mu)$-plane that separates the different cases of the spectrum of the equilibrium $\mathbf{0}$. This parabola is shown in Fig.~\ref{fig:basic}(a) as a purple curve that separates the $(\beta_{2}, \mu)$-plane into three regions, where the eigenvalues of $\mathbf{0}$ are of the form (I), (II) and (III), respectively. Note that $\mathbf{0}$ is a real saddle in region (I) and a saddle-focus in region (II). Due to the real eigenvalues of $\mathbf{0}$, homoclinic solutions in region (I) must have non-oscillatory exponentially decaying tails. In contrast, homoclinic solutions in region (II) must have oscillatory decaying tails because the eigenvalues of $\mathbf{0}$ are complex conjugates. In the following sections up to Sec.~V, we present our results on the existence of homoclinic solutions for the horizontal line $\mu=1$ in $(\beta_{2}, \mu)$-plane. This choice of $\mu$-value does not restrict the generality of our results as we will show in Sec.~V; indeed, our results extend throughout the upper half of the $(\beta_2,\mu)$-parameter plane, that is, when $\mu>0$. Figure~\ref{fig:basic}(b) and (c) show two homoclinic solutions along this horizontal line for two distinct $\beta_2$-values in region (I) and (II), respectively. The homoclinic solution in Fig.~\ref{fig:basic}(b) has non-oscillating exponentially decaying tails as it belongs to region (I). Furthermore, this homoclinic solution persist until  $\beta_2$ reachs the right-hand side of the parabola in Fig.~\ref{fig:basic}(a), that is, at the boundary between regions (II) and (III) -- which correspond to a Hamiltonian-Hopf (HH) bifurcation \cite{haragus2010local, iooss1993perturbed}.  In Fig.~\ref{fig:basic}(c), we show what this homoclinic solution looks like at $\beta_2=0.4$ where it has oscillating tails. In particular, this homoclinic solution is $R_1$-symmetric and we refer to it as the primary homoclinic solution; its corresponding soliton is the one Tam \textit{et al} considered in \cite{tam2018solitary}. Note that the homoclinic solution in Fig.~\ref{fig:basic}(c) has oscillatory decaying tails but the oscillations damp out very quickly. The oscillations in the tails of the these homoclinic solutions increase when moving horizontally towards HH bifurcation. Because the primary homoclinic solution exists at the transition between regions~(I) and (II), the conditions for a BD bifurcation are satisfied and infinitely many homoclinic solutions must exist in the indicated grey shaded region near the branch BD in Fig.~\ref{fig:basic}(a)\cite{champneys1993bifurcation}.

\section{NUMERICAL IDENTIFICATION AND CONTINUATION OF HOMOCLINIC SOLUTIONS}
\label{sec:bvp}

The task is now to find and identify a representative number of symmetric and non-symmetric homoclinic solutions of system \eqref{system}. To this end, we make use of continuation algorithms for two-point boundary value problems (2PBVP), implemented in the software package \textsc{Auto-07p} \cite{doedel2007auto} and its extension \textsc{HomCont} \cite{champneys1996numerical}. In the 2PBVP formulation, time is rescaled to the interval $[0,1]$. Thus, the integration time is treated as a free parameter that multiplies the right-hand side of system~(4), that is, 
\begin{equation}
\frac{d\mathbf{v}}{dt}=T f(\mathbf{v},\zeta).
\label{rescaled1}
\end{equation}
Note that we always assume that $T>0$. Suitable boundary conditions are imposed at the starting point $\mathbf{v}(0)$ and the end point $\mathbf{v}(1)$ of the solution segment \cite{krauskopf2007numerical}. For the continuation of a homoclinic solution, we use projection boundary conditions that place $\mathbf{v}(0)$ in the unstable eigenspace $E_{u}(\mathbf{0})$ of the equilibrium $\mathbf{0}$, and $\mathbf{v}(1)$ in one of the reversibility sections $\Sigma_{1}$ or $\Sigma_{2}$. In this way, we take advantage of the reversibility of system \eqref{system} to compute only half of a symmetric homoclinic solution $\mathbf{v}(t)$. Convergence forward in time to the stable eigenspace $E_{s}(\mathbf{0})$ is guaranteed by the corresponding reversibility conditions, and the remaining part of the homoclinic solution is obtained by applying $R_{1}(\mathbf{v}(-t))$ or $R_{2}(\mathbf{v}(-t))$. Notice that this formulation is only able to continue symmetric homoclinic solutions, since an intersection with a reversibility  section is required.

For the case of non-symmetric homoclinic solutions we formulate a 2PBVP of the  entire homoclinic solution. \textsc{Auto-07P} is a general-purpose continuation package designed for generic vector fields, and particular considerations have to be taken when continuing homoclinic and periodic solutions in reversible and Hamiltonian systems \cite{galan2014continuation}. To deal with the fact that homoclinic solutions generically persist when a single parameter is varied, we follow \cite{galan2014continuation} and introduce the gradient of the conserved quantity $H$ as a perturbation of the vector field equations
\begin{equation}
\frac{d\mathbf{v}}{dt}=T f(\bf{v},\zeta)+\delta \nabla \emph{H},
\label{rescaled2}
\end{equation}
where $\delta$ is an additional continuation parameter. We then impose the boundary conditions that $\mathbf{v}(0)$ lies in $E_{u}(\mathbf{0})$ and $\mathbf{v}(1)$ lies in $E_{s}(\mathbf{0})$. We continue the solution of the overall 2PBVP in one of the parameters while allowing $\delta$ and $T$ to vary. In this setup, the new parameter $\delta$ is free but remains extremely close to zero during continuation \cite{galan2014continuation,Doedel2003, MUNOZALMARAZ20031}. Note that the 2PBVP formulation of the entire homoclinic solution can  be used to continue symmetric homoclinic solutions as well.

To find the first homoclinic solution, we make use of a numerical implementation of Lin's method \cite{krauskopf2008lin}, where we consider two orbit segments $\mathbf{v}_a(t)$, $\mathbf{v}_b(t)$ and a suitable three-dimensional hyperplane $\Sigma$. Here, $\mathbf{v}_a(0)$ and $\mathbf{v}_b(1)$ lie in $E_u(\mathbf{0})$ and $E_s(\mathbf{0})$, respectively, and $\mathbf{v}_a(1)$ and $\mathbf{v}_b(0)$ both lie in $\Sigma$. Then the signed difference (called the Lin gap) between  $\mathbf{v}_a(1)$ and  $\mathbf{v}_b(0)$, along a fixed one-dimensional direction, provides a well-defined test function whose zeros correspond to homoclinic solutions of system \eqref{system}; see \cite{krauskopf2008lin}. Once a zero is found, the associated homoclinic solution can be followed in system parameters with the previously constructed 2PBVP formulations. Lin's method allows us to compute multi-hump homoclinic solutions of different types. We remark that the \textsc{Auto} demo rev \cite{doedel2007auto}, for the GNLSE as considered in \cite{hunt1989structural}, contain a setup to continue the basic homoclinic solutions. However, we are interested in many different types of homoclinic solutions that cannot be computed from the demo, and they are all identified with Lin's method.  

Connections between the equilibrium $\mathbf{0}$ and periodic solutions, which we refer to as EtoP connections, are organising centres for the existence of homoclinic solutions under mild conditions \cite{palis2012geometric,articleCham}; hence, they are an important object to consider when studying homoclinic solutions. To compute EtoP connections, we have to find first periodic solutions of system \eqref{system} that support a connection to $\mathbf{0}$. In reversible and Hamiltonian systems, periodic solutions are not isolated in phase space for fixed parameter values \cite{galan2014continuation}. To be able to compute and continue them, we consider the perturbed system \eqref{rescaled2} and use the 2PBVP formulation for periodic solutions \cite{krauskopf2007numerical}. For the initial data of the formulation, we use homoclinic solutions previously constructed with the 2PBVP above, as they are good initial approximation of periodic solutions of high period. Performing a continuation step in $\delta$ and $T$ allows us to find the family of periodic solutions for fixed parameter values, while $\delta$ again remains practically 0. These solution families form two-dimensional surfaces in phase space where each periodic solution lies in a particular energy level. Among these solution families, we focus on saddle periodic solutions in the zero energy level because they are the only periodic solutions that can have connections with $\mathbf{0}$. 

To find connections from $\mathbf{0}$ to a saddle periodic solution we can follow the approach that we used to find homoclinic solutions to $\mathbf{0}$. This requires one to compute first the Floquet multipliers and Floquet bundles of the periodic solution, as they contain the linear information of the flow near the periodic solution. Saddle periodic solutions of system \eqref{system} have one stable (inside the unit circle) and one unstable (outside of the unit circle) Floquet multipliers with associated stable and unstable Floquet bundles, respectively. The stable (unstable) bundle consists of the directions in phase space, along which solutions converge forward (backward) in time to the periodic solution. Finally, the other two Floquet multipliers of a saddle-periodic solution in system~\eqref{system} are always equal to one. Associated to them, there are two Floquet bundles: one that is pointing in the direction of the flow along the periodic solution, called the trivial bundle, and another one tangent to the surface of the periodic solutions in phase space. 

We compute the Floquet multipliers and their corresponding bundles with a homotopy step for a suitable 2PBVP formulation; see \cite{doi:10.1137/05062408X} for more details. Note that any connection that converges backward in time to $\mathbf{0}$ and forward in time to a $R_1$-symmetric ($R_2$-symmetric) periodic solution has a $R_1$-counterpart ($R_2$-counterpart) that converges backward in time to the same periodic solution and forward in time to $\mathbf{0}$. As we are interested in periodic solutions that are $R_1$-symmetric or $R_2$-symmetric, we make use of this fact to set up Lin's method by using the unstable eigenspace $E_u(\mathbf{0})$ of $\mathbf{0}$ and the stable bundle of the periodic solution. That is, we consider two orbits segments $\mathbf{v}_a(t), \mathbf{v}_b(t)$ and a suitable three-dimensional hyper plane $\Sigma$, such that, $\mathbf{v}_a(0)$ and $\mathbf{v}_b(1)$ lie in $E_u(\mathbf{0})$ and the stable Floquet bundle of the saddle periodic solution, respectively, and $\mathbf{v}_a(1)$ and $\mathbf{v}_b(0)$ lie in $\Sigma$. The zeros of the corresponding Lin gap correspond to EtoP connections that converge backwards in time to $\mathbf{0}$ and forward in time to the periodic solution.  

\section{HOMOCLINIC FAMILIES OF DIFFERENT TYPES }
\label{sec:connorbits} 

Computing EtoP connections between $\mathbf{0}$ and saddle periodic solutions in the zero-energy level is a good starting point for understanding how different homoclinic solutions are organised. Since system \eqref{system} is reversible, existence of a connection from $\mathbf{0}$ to a \emph{symmetric} saddle periodic solution guarantees a return connection from the periodic solutions to $\mathbf{0}$ as well. This return connection corresponds to the $R_1$- or $R_2$-counterpart of the EtoP connection, depending on the symmetry of the periodic solution. The existence of a connection from $\mathbf{0}$ to a periodic solution and a connection back from the periodic solution to $\mathbf{0}$ is known as a heteroclinic cycle. Existence of these heteroclinic cycles, and their persistence under parameter variation (transversality), generate a mechanism for the existence of homoclinic solutions that go around the periodic solution multiple times, as a consequence of the $\lambda$-lemma \cite{wiggins2003introduction, palis2012geometric}. Thus, it is possible to find homoclinic solutions that   
\begin{itemize}
\item[(a)]{follow closely an EtoP connection from $\mathbf{0}$ to the periodic solution,}
\item[(b)]{then loop $n$-times close to the periodic solution, and}
\item[(c)]{follow closely an EtoP connection from the periodic solution back to $\mathbf{0}$. }
\end{itemize}

The different combinations of EtoP connections generate cycles with different symmetry properties that organise specific homoclinic solution families in parameter space. Some of these families are organised by EtoP connections to a $R^*$-symmetric periodic solution, and others by EtoP connections to a $R_1$-symmetric periodic solution. The corresponding solitons associated with these homoclinic families are distinct from bound states of two or more primary solitons since the spacing of the maxima is fixed and different from the location of the zeros of the oscillating tails. In what follows, we study different families of such homoclinic solutions. We show bifurcation diagrams in $\beta_2$ of these families to illustrate how they persist and coalesce due to the existence of different underlying EtoP connections.

\subsection{HOMOCLINIC SOLUTIONS ASSOCIATED WITH $R^*$-SYMMETRIC PERIODIC SOLUTION}
\label{sec:RstarPeriodic}

\begin{figure*}[t!]
   \centering
   \includegraphics[width=\textwidth]{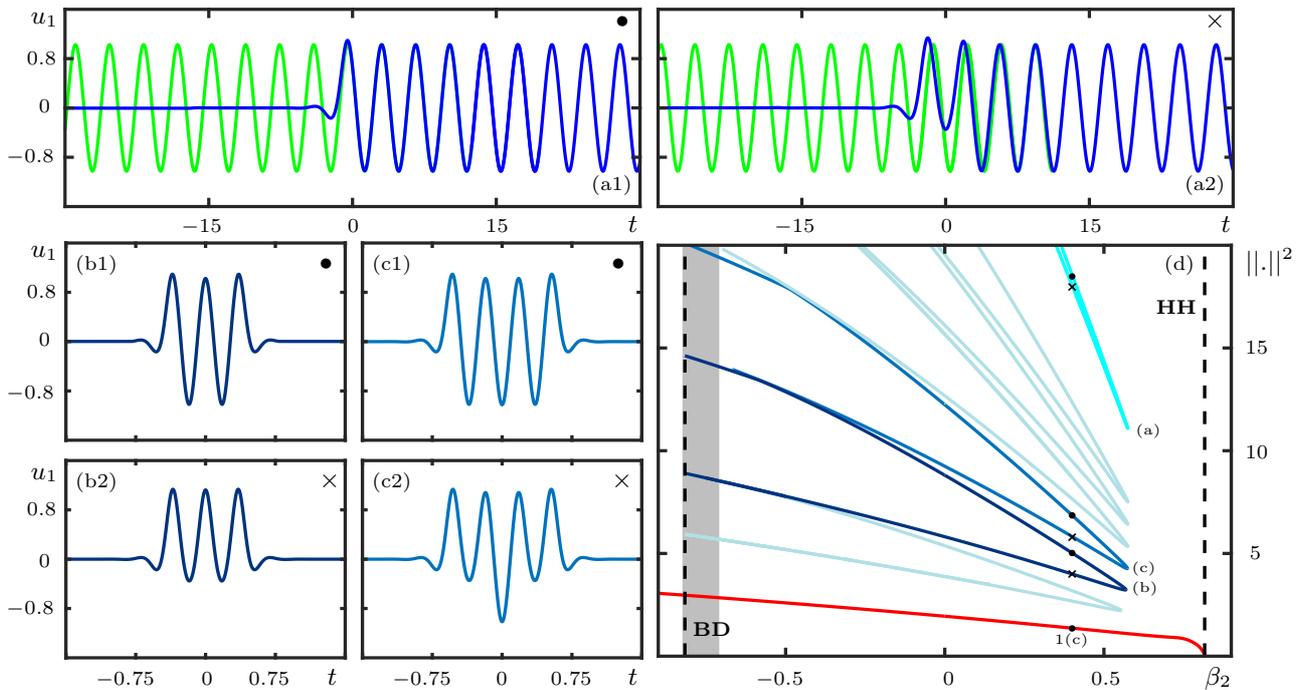}
    \caption{Family of $R_1$-symmetric homoclinic solutions associated with $R^*$-symmetric periodic solution. Panels~(a1)-(a2) show two EtoP connections (blue curve) between $\mathbf{0}$ and a periodic solution $\Gamma_{*}$ (green curve) that is invariant under both $R_{1}$ and $R_{2}$. Panels~(b1),(c1) and (b2),(c2) show temporal traces of $R_{1}$-symmetric homoclinic solutions, associated with the connections shown in panels~(a1) and (a2), respectively. Panel (d) shows the bifurcation diagram in $\beta_2$ of the EtoP connections (cyan) and $R_{1}$-symmetric homoclinic solutions, where solutions are represented by the square of the $L_2$-norm of their $u_1$ component, and each $R_{1}$-symmetric homoclinic curve represents a family of homoclinic solutions that have the same number of humps. Notice that the colour of a homoclinic solutions in panels~(b) and (c) and their corresponding bifurcation curve in panel (d) is the same. The black dot on the red curve corresponds to the primary homoclinic solution shown in Fig.~\ref{fig:basic}(c); the black dot and black cross on other bifurcation curves correspond to the solutions shown in panels (1) and (2), respectively. The black dashed lines delimit the parameter interval where $\mathbf{0}$ has complex eigenvalues with non-zero real parts; the one on the left indicates the BD bifurcation, and the one on the right the HH bifurcation. The shaded grey region represents the region close to the purple colour parabola in Fig.~\ref{fig:basic}(a). Also shown are the bifurcation curves (light-blue) of the $R_{1}$-symmetric homoclinic solutions that have two, five, six and seven humps. The bifurcation curves in panels (d) are for $(\beta_{4}, \gamma, \mu)=( -1, 1, 1)$; moreover, $\beta_2=0.4$ in panels (a1)-(c2).} 
\label{fig:R1symm}
\end{figure*} 

\begin{figure*}[t!]
   \centering
   \includegraphics[width=\textwidth]{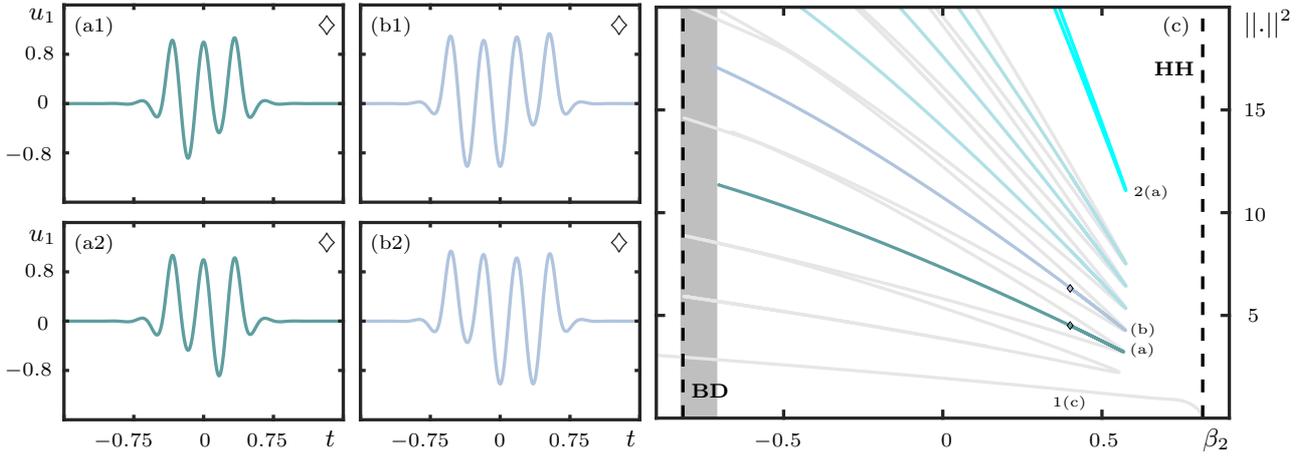}
    \caption{Family of $R_1$-symmetry broken homoclinic solutions associated with $\Gamma_{*}$ . Panels (a) and (b) show temporal traces of non-symmetric homoclinic solutions associated with the connections shown in Fig.~\ref{fig:R1symm}(a). Panel (c) shows the bifurcation diagram in $\beta_2$ of the EtoP connections (cyan) and $R_1$-symmetry broken homoclinic solutions  where solutions are represented by the square of the $L_2$-norm of their $u_1$ component. The black diamonds on the bifurcation curves correspond to the solutions shown in panels (a1)-(b2). Panel~(c) follows the same colour and symbols convention as Fig.~\ref{fig:R1symm}(d) but with respect the $R_1$-symmetry broken homoclinic solutions; all the homoclinic bifurcation curves from Fig.~\ref{fig:R1symm}(d) are superimposed in light grey in panel~(c). The bifurcation curves in panels (c) are for $(\beta_{4}, \gamma, \mu)=( -1, 1, 1)$; moreover, $\beta_2=0.4$ in panels (a1)-(b2).} 
    \label{fig:R1symmbroken}
\end{figure*} 

Heteroclinic cycles from $\mathbf{0}$ to a $R^*$-symmetric periodic solution organise two main families of homoclinic solutions: the $R_1$- and $R_2$-symmetric families. Furthermore, these families also organise non-symmetric homoclinic solutions that arise when the corresponding reversibility condition is broken. 

\subsubsection{$R_1$-symmetric homoclinic solutions }  

Figure~\ref{fig:R1symm} illustrates a family of solutions associated with a $R^*$-symmetric periodic solution $\Gamma_{*}$ along with their bifurcation diagram. Panels (a1) and (a2) show two distinct EtoP connections (blue) from $\mathbf{0}$ to $\Gamma_{*}$ (green) for $(\beta_{2},\beta_{4},\gamma,\mu)=(0.4,-1,1,1)$. Both these EtoP connections converge backward in time to $\mathbf{0}$ and forward in time to $\Gamma_{*}$; however,  these connections are not related by symmetry as their temporal profiles are different and cannot be mapped to each other by any of the reversibilities or the spatial-temporal symmetry. In particular, the EtoP connection shown in panel~(a1) makes a small negative excursion in $u_1$ and then has a transient for positive $u_1$ before converging to $\Gamma_{*}$ after $t \approx 0$; in panel~(a2), on the other hand, it makes two oscillation for positive $u_1$ before it traces $\Gamma_{*}$ from $t \approx 4$. Since $\Gamma_{*}$ is $R^*$-symmetric, there exist the $R_1$- and $R_2$-counterparts of the EtoP connections, which are reflections in $t$ and rotations by $180^{\degree}$ of panels (a), respectively. The different combinations of these EtoP connections create different heteroclinic cycles that organise different types of homoclinic solutions. 

We first consider the heteroclinic cycle that is formed by the EtoP connection shown in panel (a1) and its corresponding $R_1$-counterpart. The temporal profiles of the $u_1$-component of two homoclinic solutions organised by this cycle are shown in Fig.~\ref{fig:R1symm}(b1) and (c1). The homoclinic solution in panel (b1) can be thought of as a concatenation of first part of the EtoP connection in panel (a1) up to its second maximum with its $R_1$-symmetric counterpart. Similarly, the homoclinic solution in panel (b2) is associated with the EtoP cycle formed by the connection in panel (a2) and its $R_1$-symmetric counterpart. Two further homoclinic solutions are shown in panels (c1) and (c2); they are derived from the EtoP connections in panels (a1) and (a2) in the same way but for one further half-turn around $\Gamma_{*}$. 

This type of homoclinic solutions exist for any number of humps, including those with three and four shown in panels (b1)-(c2). They all exhibit oscillating tails with oscillations that damp out quickly, and these homoclinic solutions follow the respective EtoP connection in panels (a) to intersect $\Sigma_1$ transversally at $t=0$ where they start following the $R_1$-symmetric counterpart. Hence, they are $R_1$-symmetric homoclinic solutions. Note that the $R_2$-counterparts of the $R_1$-symmetric homoclinic solution also exist. However, we do not show them here because, on the level of this figure, they correspond to reflections of $u_1$ in the $t$-axis so that maxima become minima and vice-versa. 

Figure~\ref{fig:R1symm}(d) shows the bifurcation diagram of the EtoP connections and $R_1$-symmetric homoclinic solutions in the $(\beta_{2},||u_{1}||^2)$-plane. Here, the dotted vertical lines bound the interval ($-0.8164,0.8164$), where the equilibrium $\mathbf{0}$ is a saddle-focus; hence, this interval represents the $\beta_2$-values, between BD and HH, where homoclinic solutions with oscillating tails exist; the shaded grey region represents the region close to the purple parabola in Fig.~\ref{fig:basic}(a). 
The EtoP connections in Fig.~\ref{fig:basic}(a) lie on a single curve with two branches that meet at a fold at $\beta_{2}\approx0.5753$; the EtoP connections in panels (a1) and (a2) are from the upper and the lower branch of this curve, respectively. As the EtoP connections have an infinite $L_2$-norm, we represent them in panel~(d) with a finite norm by truncating the connection after ten oscillations near the periodic solution. The parameter value where they fold is the moment where two $R_1$-symmetric EtoP cycles coalesce; they no longer exist beyond that value. Hence, pairs of $R_1$-symmetric EtoP 
cycles exist for $\beta_2 \in (-0.8164,0.5753)$ and they come together at $\beta_{2}\approx0.5753$. This has far-reaching consequences for the organisation of the two families of $R_1$-symmetric homoclinic solutions associated with $\Gamma_{*}$, as is illustrated in panel~(d). All $R_1$-symmetric homoclinic solutions also lie on curves with two branches that meet at fold points, where two $R_1$-symmetric homoclinic solutions coalesce. For each curve, the upper branch corresponds to the homoclinic solutions associated with the EtoP cycle generated by the connection in panel (a1) and its $R_1$-counterpart, while the lower branch corresponds to the homoclinic solutions associated with the EtoP cycle generated by the connection in panel (a2) and its $R_1$-counterpart. In Fig.~\ref{fig:R1symm}(d), we show the bifurcation curves of the homoclinic solutions with two to eight humps; the two curves that are highlighted in darker colour correspond to the homoclinic solutions with three and four humps shown in panels (b1)-(c2). Notice that all the bifurcation curves associated with the $R_1$-symmetric homoclinic solutions of $\Gamma_{*}$ fold close to $\beta_{2}\approx0.5753$. Furthermore, as the number of humps of the homoclinic solutions increases, the $\beta_2$-values where they fold approach $\beta_{2}\approx0.5753$ from below; that is, they accumulates on the $\beta_2$-values where the EtoP connection folds. Also shown in panel (d) is the curve of the primary homoclinic solution from Fig.~\ref{fig:basic}(c), which exists in the entire $\beta_2$-range up to $\beta_{2}\approx0.8164$ where the eigenvalues of $\mathbf{0}$ becomes purely imaginary at HH.

\subsubsection{$R_1$-symmetry broken homoclinic solutions }

Since there exist two distinct EtoP connections to $\Gamma_{*}$, one can also consider the heteroclinic cycle that is formed by the EtoP connection in Fig.~\ref{fig:R1symm}(a1) and the $R_1$-counterpart of the EtoP connection in Fig.~\ref{fig:R1symm}(a2), or vice-versa. The homoclinic solutions associated with these cycles are non-symmetric and are illustrated in Fig.~\ref{fig:R1symmbroken} along with their bifurcation diagram. The homoclinic solution in panel (a1) can be thought of as a concatenation of the first part of the EtoP connection in Fig.~\ref{fig:R1symm}(a1) up to its second maximum with the $R_1$-counterpart of the EtoP connection in Fig.~\ref{fig:R1symm}(a2) up to its second maximum. If the concatenation is performed the other way around, the homoclinic solution in panel (a2) is obtained; it is the $R_1$-counterpart of the homoclinic solution in panel (a1). By considering one further half-turn around $\Gamma_{*}$, the homoclinic solutions in panels (b1) and (b2) are derived. In this way, non-symmetric homoclinic solutions for any number of humps can be obtained.

Note that all these non-symmetric homoclinic solutions also come in pairs, but they are each others $R_1$-counterparts. Hence, in the bifurcation diagram in Fig.~\ref{fig:R1symmbroken}~(c) the two branches lie on top of each other and are indistinguishable. The two branches meet at a fold point at $\beta_{2}\approx0.5753$, and they become $R_1$-symmetric at this point. As before, we show the bifurcation curves of the EtoP connection and non-symmetric homoclinic solutions from three to seven humps in panel (c). In particular, we highlight the bifurcation curves of the three- and four-hump non-symmetric homoclinic solutions in a darker colour; moreover, all curves of the $R_1$-symmetric homoclinic bifurcations from Fig.~\ref{fig:R1symm}(d) are shown in light-grey. Note that non-symmetric and $R_1$-symmetric homoclinic solutions with the same number of humps fold at the same $\beta_2$ value. That is, in order to transition between corresponding $R_{1}$-counterparts of each non-symmetric homoclinic solution, they must reach a fold point where they become symmetric. Thus, each fold point is a symmetry-breaking of the $R_1$ symmetry. Therefore, we refer to this family of non-symmetric homoclinic solutions as $R_1$-symmetry broken homoclinic solutions of $\Gamma_{*}$. 

\subsubsection{$R_2$-symmetric and $R_2$-symmetry broken homoclinic solutions }

There also exit cycles formed by the EtoP connections shown in Fig.~\ref{fig:R1symm}(a) and their $R_2$-counterparts. In general, we find a similar phenomenon where the corresponding cycle organises $R_2$-symmetric homoclinic solutions which come in pairs. In particular, these homoclinic solutions intersect the reversibility section $\Sigma_2$ transversally at $t=0$.  They symmetry break at fold points and there are also associated pairs of non-symmetric homoclinic solutions, which are $R_2$-symmetry broken solutions.   

\begin{figure*}[t!]
   \centering
   \includegraphics[width=\textwidth]{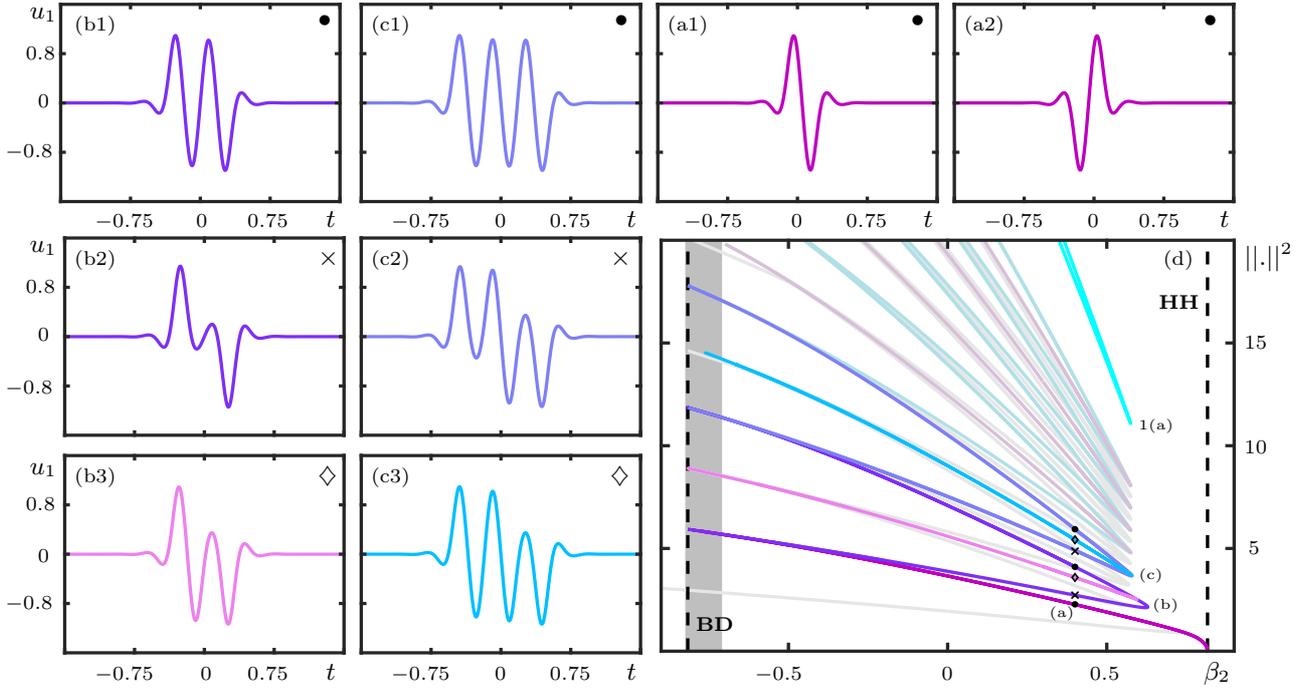}
    \caption{Family of $R_2$-symmetric and $R_2$-symmetry broken homoclinic solutions associated with $\Gamma_{*}$. Panels (a) show the temporal traces of the basic $R_2$-symmetric homoclinic solution and its corresponding $R_1$-symmetric counterparts. Panels (b1),(c1) and (c2),(c3) show the temporal traces of the $R_2$-symmetric homoclinic solutions associated with the EtoP connections shown in Fig.~\ref{fig:R1symm}(a1) and Fig.~\ref{fig:R1symm}(a2), respectively.  Panels (b2) and (c2) show non-symmetric homoclinic solutions with one and two humps, respectively. Panel (d) shows the bifurcation diagram in $\beta_2$ of the EtoP connections (cyan) and corresponding homoclinic solutions, where solutions are represented by the square of the $L_2$-norm of their $u_1$ component. Panel~(d) follows the colour and symbols convention as Fig.~\ref{fig:R1symm}(d) but with respect the $R_2$-symmetric and $R_2$-symmetry broken homoclinic solutions; all the homoclinic bifurcation curves from Fig.~\ref{fig:R1symm}(d) and Fig.~\ref{fig:R1symmbroken}(c) are superimposed in light grey in panel~(d). The bifurcation curves in panels (d) are for $(\beta_{4}, \gamma, \mu)=( -1, 1, 1)$; moreover, $\beta_2=0.4$ in panels (a1)-(c3).} 
\label{fig:R2symm}
\end{figure*} 

Figure~\ref{fig:R2symm} shows some representative examples of $R_2$-symmetric and $R_2$-symmetry broken homoclinic solutions together with their bifurcation diagram. As for the $R_1$-symmetric homoclinic solutions, there exists a basic $R_2$-symmetric homoclinic solution with one hump, which is shown in panel (a1). To illustrate the effect of the $R_1$-reversibility on these $R_2$-symmetric homoclinic solutions, the $R_1$-counterpart of this solution is shown in panel (a2). The homoclinic solutions in panels (b1)-(b2) and (c1)-(c2) have one and two further half-turns around $\Gamma_{*}$, respectively. Two non-symmetric homoclinic solutions are shown in panels (b3) and (c3); they are associatd with the EtoP cycle formed by the EtoP connections in Fig.~\ref{fig:R1symm}(a1) and the $R_2$-counterpart of the EtoP connection in Fig.~\ref{fig:R1symm}(a2).  

Figure~\ref{fig:R2symm}(d) shows the bifurcation diagram of the $R_2$-symmetric and non-symmetric homoclinic solutions up to seven humps, and we highlight the ones with two and three humps in a darker colour. Here, all the previously shown homoclinic bifurcation curves are shown in light grey. The basic $R_2$-symmetric homoclinic solution and its $R_1$-counterpart shown in panels (a), exist throughout the $\beta_2$-interval where $\mathbf{0}$ has complex conjugate eigenvalues. On the other hand, all the other $R_2$-symmetric homoclinic solutions lie again on curves with two branches that meet at fold points. The new non-symmetric homoclinic solutions also come in pairs. Since they have the same $L_2$-norm, the respective two branches of the bifurcation curves lie on top of each other.  All non-symmetric homoclinic solutions become $R_2$-symmetric at the coinciding fold points; here the $R_2$-symmetry is broken, which is why we refer to them as $R_2$-symmetry broken homoclinic solutions. As the number of humps per homoclinic solution increases, the parameter values where they fold accumulate on $\beta_{2}\approx0.5753$; notice however, that this accumulation is now from larger values of $\beta_2 $. 

\begin{figure*}[t!]
   \centering
   \includegraphics[width=\textwidth]{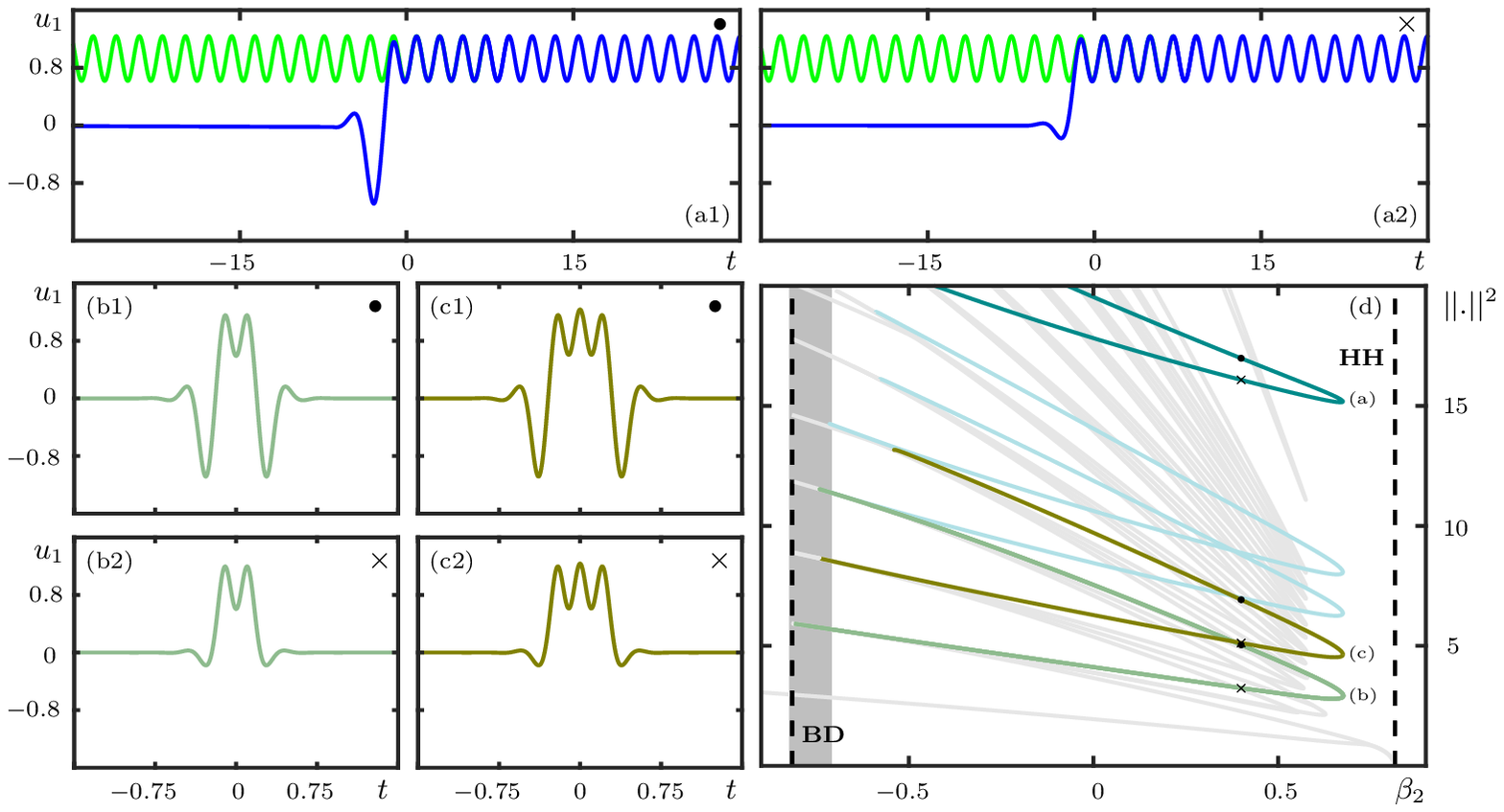}
    \caption{Family of $R_1$-symmetric homoclinic solutions associated with $R_1$-symmetric periodic solution. Panels~(a1) and (a2) show two connections (blue curve) between $\mathbf{0}$ and a periodic solution $\Gamma_{1}^{+}$ (green curve) that is only invariant under $R_{1}$. Panels~(b1),(c1) and (b2),(c2) show temporal traces of $R_{1}$-symmetric homoclinic solutions, associated with the connections shown in panels~(a1) and (a2), respectively. Panel (d) shows the bifurcation diagram in $\beta_2$ of the EtoP connections (cyan) and $R_{1}$-symmetric homoclinic solutions, where solutions are represented by the square of the $L_2$-norm of their $u_1$ component. All the bifurcation curves  from Fig~\ref{fig:R1symm}(d), Fig.~\ref{fig:R1symmbroken}(c) and Fig.~\ref{fig:R2symm}(d) are superimposed in light grey.  Panel (d) follows the colour and symbol convention as Fig.~\ref{fig:R1symm}(d) but with respect to the $R_1$-symmetric homoclinic solutions associated with $R_1$-symmetric periodic solution. The bifurcation curves in panels (d) are for $(\beta_{4}, \gamma, \mu)=( -1, 1, 1)$; moreover, $\beta_2=0.4$ in panels (a1)-(b2). } 
\label{fig:R1symmR1}
\end{figure*} 

\subsubsection{Connection with homoclinic snaking}

It is clear from Fig.~\ref{fig:R2symm}(d) that only the bifurcation curves of the primary $R_1$- and $R_2$-symmetric homoclinic solutions reach the HH bifurcation; the bifurcation curves of multi-hump $R_1$- and $R_2$-symmetric homoclinic solutions, on the other hand, have folds before reaching HH. When viewed for decreasing $\beta_2$, the two branches of primary homoclinic solutions emerge from the HH bifurcation. It has been observed in other four-dimensional reversible systems \cite{woods1999heteroclinic}, including the Swift-Hohenberg equation \cite{articleBruke} and the Lugiato-Lefever equation (LLE) \cite{articleParra}, that these primary homoclinic curves born at the HH bifurcation can undergo a phenomenon known as \emph{homoclinic snaking}: these two branches of homoclinic solutions fold back and forth repeatedly when continued in a chosen parameter. Moreover, there exist branches of symmetry-broken homoclinic solutions that connect the two branches of symmetric homoclinic solutions at respective fold points; these symmetry broken branches are also referred to as \enquote{rungs} because they form a ladder-like structure with the two primary branches.

The bifurcation structure we find here for system~\eqref{system} in Figs.~\ref{fig:R1symm}--\ref{fig:R2symm} is quite similar in spirit, but the bifurcation curves of all homoclinic solutions end for decreasing $\beta_2$ at the BD bifurcation rather than featuring fold bifurcation on the left as well. This type of bifurcation structure due to the existence of the BD bifurcation, which we refer to as \emph{BD-truncated homoclinic snaking}, was observed, for example, in \cite{articleParra} in a certain parameter regime of the LLE. In contrast to the LLE, changing any of the parameters of system \eqref{system} does not qualitatively change the bifurcation diagram in Fig.~\ref{fig:R2symm}(d), as can be seen from the non-dimensionalisation. Thus, one cannot find full homoclinic snaking in system~\eqref{system}. 

The absence of homoclinic snaking means, in particular, that the branches of symmetric homoclinic solutions with increasing numbers of humps of system \eqref{system} do not form two single connected branches. Hence, they cannot be obtained simply by continuation of the two primary homoclinic solutions through successive fold points but must be found one-by-one. As was explained in Sec.~\ref{sec:bvp}, this can be achieved efficiently with Lin's method. This approach has the additional advantage that it allows us to also find and continue the underlying EtoP connections, which organise the respective branches of homoclinic solutions with different symmetry properties. Finding branches of EtoP connections is a new aspect of our work, which shows that the $\beta_2$-values of the fold points of homoclinic solution curves accumulate, as the number of humps increases, on the $\beta_2$-value of the fold of the underlying EtoP connection. As we will show next, there are more such EtoP connections, including those to periodic solutions with less symmetry.

\subsection{HOMOCLINIC SOLUTIONS ASSOCIATED WITH $R_1$-SYMMETRIC PERIODIC SOLUTION}
\label{sec:R1Periodic}

\begin{figure*}[t!]
   \centering
   \includegraphics[width=\textwidth]{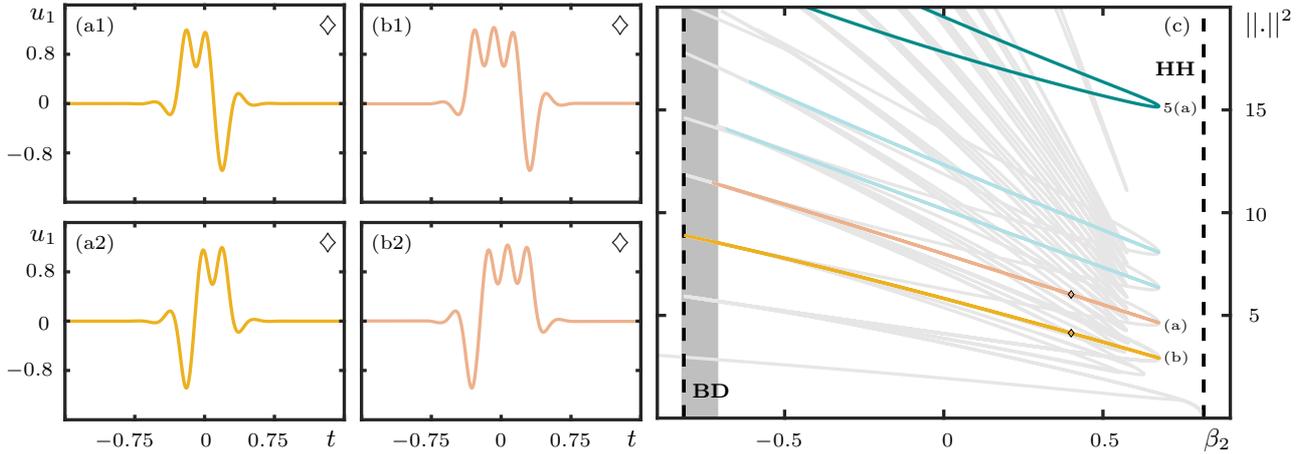}
    \caption{Family of $R_1$-symmetry broken homoclinic solutions associated with $\Gamma_{1}^{+}$. Panels (a) and (b) show temporal traces of non-symmetric homoclinic solutions associated with the connections shown in Fig.~\ref{fig:R1symmR1}(a). Panel (c) shows the bifurcation diagram in $\beta_2$ of the EtoP connections (cyan) and $R_1$-symmetry broken homoclinic solutions where solutions are represented by the square of the $L_2$-norm of their $u_1$ component. Panel~(c) follows the colour and symbols convention as Fig.~\ref{fig:R1symm}(d) but with respect to the $R_1$-symmetry broken homoclinic solutions associated with $R_1$-symmetric periodic solution; furthermore, all the previously shown homoclinic bifurcation curves are superimposed in light grey in panel~(c). The bifurcation curves in panels (d) are for $(\beta_{4}, \gamma, \mu)=( -1, 1, 1)$; moreover, $\beta_2=0.4$ in panels (a1)-(b2).} 
\label{fig:R1symmbrokenR1}
\end{figure*}

It is possible to have a (pair of) periodic solutions with only $R_1$-symmetry in the zero-energy level. We find that there are EtoP connections between $\mathbf{0}$ and these periodic solutions for certain parameter values. As before, there are associated $R_1$-symmetric homoclinic solutions that come in pairs and meet at fold points, where they also symmetry break. However, we do not find $R_2$-symmetric and $R_2$-symmetry broken homoclinic solutions associated with these EtoP connections. Families of $R_1$-symmetric and $R_1$-symmetry broken homoclinic solutions associated with an $R_1$-symmetric periodic solution are shown respectively in Fig.~\ref{fig:R1symmR1} and~\ref{fig:R1symmbrokenR1}. 

Figure~\ref{fig:R1symmR1} shows the $R_1$-symmetric homoclinic solutions in the same layout as Fig.~\ref{fig:R1symm}. Panels (a1) and (a2) show two EtoP connections, but now to an $R_1$-symmetric periodic solution $\Gamma_{1}^{+}$. Note that throughout this manuscript, the $R_2$-counterpart of $\Gamma^+$ is denoted by $\Gamma^-$, and that any results pertaining to homoclinic solutions and EtoP connections of $\Gamma^+$ also applies to $\Gamma^-$. The EtoP connections in panels (a) are not related by symmetry: the one in panel~(a1) has a larger negative excursion in $u_1$ before converging to $\Gamma_{1}^{+}$ compared to that in panel~(a2). Associated with the EtoP cycles generated by the EtoP connections in panels (a) and their corresponding $R_1$-counterparts, one can find $R_1$-symmetric homoclinic solutions that make any number of turns around $\Gamma_{1}^{+}$. Figure~\ref{fig:R1symmR1}(b1) and (c1) show homoclinic solutions with one full turn around $\Gamma_{1}^{+}$, and panels (b2) and (c2) those with one further half-turn around $\Gamma_{1}^{+}$.  Figure~\ref{fig:R1symmR1}(d) shows the corresponding bifurcation diagram, where the two EtoP connections occur on a branch that folds at $\beta_{2}\approx0.6756$; they are again represented by a finite norm (by truncating them after eight oscillations around the periodic solution). Also shown are curves of the $R_1$-symmetric homoclinic solutions from two to five humps, where the ones in panels (b)-(c) are highlighted in a darker colour. The curves of $R_1$-symmetric homoclinic solutions all have folds and, as the number of humps increases, the $\beta_2$-values where they fold accumulate onto that of the fold of EtoP connections of $\Gamma_{1}^{+}$. 

Figure~\ref{fig:R1symmbrokenR1} illustrates the $R_1$-symmetry broken homoclinic solutions, which are are associated with the EtoP cycle generatedd by the EtoP connection in Fig.~\ref{fig:R1symmR1}(a1) and the $R_1$-counterpart of the EtoP connection in Fig.~\ref{fig:R1symmR1}(a2), or vice versa. In Fig.~\ref{fig:R1symmbrokenR1}(a) these homoclinic solutions have one turn around $\Gamma_{1}^{+}$, while in panels (b) they make one further half-turn. Moreover, the homoclinic solutions in panels (a2) and (b2) are the $R_1$-counterparts of those in panels (a1) and (b1). As the bifurcation diagram in Fig.~\ref{fig:R1symmbrokenR1}(c) shows, the $R_1$-symmetry broken homoclinic connections can be found along curves that have folds at the fold points on the curves of $R_1$-symmetric homoclinic connections (light grey). Here, we highlight the curves of the $R_1$-symmetry broken homoclinic solutions with two and three humps in a darker colour; the respective two branches are again indistinguishable in panel~(c) because they have the same $L_2$-norm. 

All these curves in Fig.~\ref{fig:R1symmR1}(d) and Fig.~\ref{fig:R1symmbrokenR1}(c) of homoclinic connections associated with $\Gamma_{1}^{+}$ extend on the left to the BD bifurcation and, therefore, constitute a further instance of BD-truncated homoclinic snaking. In contrast to the bifurcation curves of the homoclinic solutions associated with $\Gamma_{*}$, there do not exist two primary homoclinic bifurcation curves that arise from the HH point. Notice also that the homoclinic solutions associated with $\Gamma_{1}^{+}$ exist over a larger $\beta_2$-interval; this is due to the fact that the fold of the bifurcation curve of EtoP connection to $\Gamma_{1}^{+}$ has a considerably larger $\beta_2$-value than the fold of the bifurcation curve of EtoP connection to the periodic solution $\Gamma_{*}$.

\section{EXISTENCE OF SOLITONS IN TWO-PARAMETER PLANEs}
\label{sec:parplane}

\begin{figure}[t!]
\begin{center}
\includegraphics[scale=1.2]{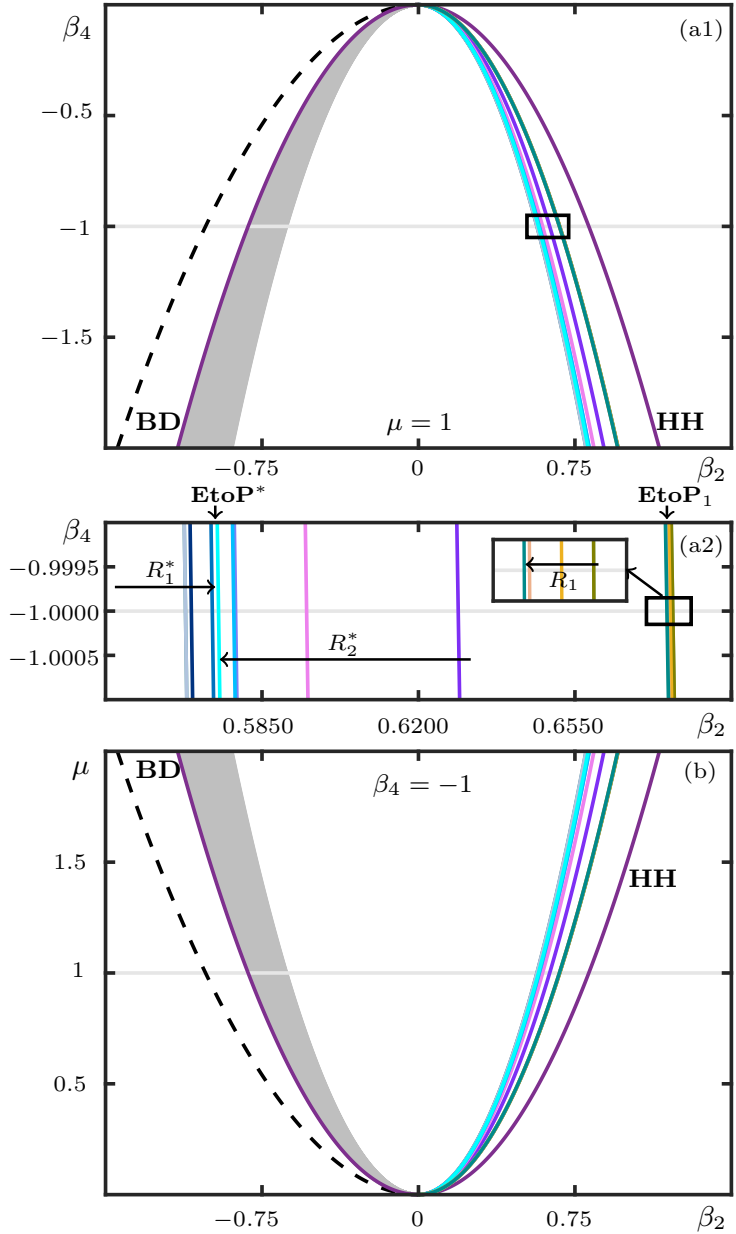}
\caption{Bifurcation diagrams in the $(\beta_2, \beta_4)$-plane for $(\mu,\gamma)=(1,1)$ in panel (a) and in the $(\beta_2, \mu)$-plane for $(\beta_{4},\gamma)=(1,1)$ in panel (b). Shown are curves of the bifurcations BD and HH (purple parabola) together with curves of folds of the identified EtoP connections and homoclinic solutions (colors as in previous figures); the black dashed curve represents the family of Karlsson-H\"{o}\"{o}k solutions. Panel (a2) is an enlargement of the rectangular region in panel (a1); it illustrates the accumulation process of folds of $R_1$- and $R_2$-symmetric homoclinic solutions onto the fold of the EtoP connection to $\Gamma_{*}$, as well as of the $R_1$-symmetric homoclinic solutions onto the fold of the EtoP connection to $\Gamma_{1}^{+}$ (enlarged further in the inset).}
\label{fig:parplane}
\end{center}
\end{figure}

As we have seen in the previous sections for fixed $(\beta_4, \mu, \gamma)=(-1,1,1)$, infinitely many homoclinic solutions and associated EtoP connections are created at the BD bifurcation, while only the two primary homoclinic solutions reach the HH bifurcation. All other homoclinic solutions and the EtoP connections disappear at fold bifurcations. We know from Sec.~\ref{sec:analysis} that the BD and HH bifurcations occur along the left and right halfs of a parabola in the $(\beta_2,\mu)$-plane of Fig.~\ref{fig:basic}(a) and also in the $(\beta_2,\beta_4)$-plane. The folds of homoclinic solutions and EtoP connections are well defined codimension-one bifurcations that we can continue numerically as curves in these parameter planes. However, this is not necessary because, as we show now, all fold bifurcations also occur along half-parabolas in either of these planes; in particular, they do not depend on the parameter $\gamma$ and they always occur in the same order as a function of $\beta_2$. 

To see this, we consider the ansatz $U(x,\tau)=\hat{u}(\tau)e^{iqz}$ with the non-dimensionalisation of the GNLSE \eqref{gnse} for $\beta_2>0$ and $\beta_4<0$, which is case~(c) in Table~\ref{tab:reductions}, to obtain the ODE
\begin{equation}
         \frac{d^{4}\hat u}{d\tau^{4}}+ \frac{d^{2}\hat u}{d\tau^{2}}+q \hat u- \hat u^3=0,
        \label{ode1}
\end{equation}	
On the other hand, the transformation 
$$\overline{u}=\sqrt{\cfrac{-\beta_{4}\gamma}{6\beta_{2}^{2}}}u,\quad \tau_0=\sqrt{\cfrac{-12\beta_{2}}{\beta_{4}}} t$$
allows us to rewrite Eq.~\eqref{ode} as
 	\begin{equation}
         \frac{d^{4} \overline{u}}{d\tau_0^{4}}+ \frac{d^{2}\overline{u}}{d\tau_0^{2}}-\frac{\beta_4 \mu}{6\beta_2^{2}} \overline{u}- \overline{u}^3=0.
        \label{ode2}
        \end{equation}  
Direct comparison between the coefficients of Eq.~\eqref{ode1} and those of Eq.~\eqref{ode2} gives
  	\begin{equation}
	6q\beta_2^{2}+\beta_4 \mu=0.
	\label{relationship}
	\end{equation}  
This relationship extends the results of our bifurcation analysis from the previous sections for fixed $\beta_4$ and $\mu$ to the whole $(\beta_2,\beta_4,\mu)$-space. Notice that the value of $\gamma$ does not influence the location of the fold bifurcations, as this parameter only affects the amplitude of the homoclinic solution; see Table~\ref{tab:reductions}. Indeed, Eq.~\eqref{relationship} shows specifically that all curves of codimension-one bifurcations with $\beta_2>0$ and $\beta_4<0$ are half-parabolas in both the $(\beta_2,\beta_4)$-plane for fixed $\mu$ and in the $(\beta_2,\mu)$-plane for fixed $\beta_4$; the respective parabola is determined from the computed $\beta_2$-values for fixed $\beta_4$ and $\mu$ by determining the respective value of $q$ in Eq.~\eqref{relationship}.

Figure~\ref{fig:parplane} illustrates this result by showing the half-parabolas of all folds of EtoP connections and multi-hump homoclinic solutions we detected. Panel~(a1) shows the $(\beta_2,\beta_4)$-plane for $\mu=1$ and panel~(b) shows the $(\beta_2,\mu)$-plane for $\beta_4=-1$, respectively; here $\gamma = 1$. Also shown are the bifurcation curves BD and HH, as well as the half-parabola along which one finds the Karlson and H\"{o}\"{o}k solution \cite{karlsson1994soliton} (which lies to the left of BD as is concerns solitons with non-oscillating decaying tails). Notice from Fig.~\ref{fig:parplane}(a1) that, for fixed quadratic dispersion  $\beta_2 > 0$, different families of homoclinic solutions arise as the quartic dispersion $\beta_4$ is decreased, namely at definite negative threshold $\beta_4$-values given by the half-parabolas of fold bifurcations. Panel~(a2) is an enlargement of the rectangular region near $\beta_4=-1$ in panel~(a1) that illustrates how the curves of the homoclinic solutions associate with $\Gamma_{*}$ and $\Gamma_{1}^{+}$ accumulate on the two basic EtoP connections from Sec.~\ref{sec:RstarPeriodic}, labelled here $\mathbf{EtoP^{*}}$, and from Sec.~\ref{sec:R1Periodic}, labelled here $\mathbf{EtoP_{1}}$. 

As Fig.~\ref{fig:parplane} shows, the ordering of these bifurcation curves is exactly the same in the $(\beta_2,\mu)$-plane in panel~(b); compare with Fig.~\ref{fig:basic}(a). Importantly, $\mu$ is not a system parameter of the GNLSE~\eqref{gnse} but arises from the ansatz~\eqref{ansatz}. Therefore, moving along any vertical line in Fig.~\ref{fig:parplane}(b) does not change any of the dispersion terms of the GNLSE. Moroever, for given quartic dispersion $\beta_4$, there is a critical $\mu$-value for solitons to exist. Therefore, by increasing the wave number $\mu$, for given fixed values of $\beta_{2}$, $\beta_{4}$ and  $\gamma$, one can generate many more homoclinic solutions and, therefore, different solitons of \eqref{gnse}.

\section{INFINITELY MANY PERIODIC SOLUTIONS WITH ZERO ENERGY}
\label{sec:perorbits}

As the previous sections show, periodic solutions of system \eqref{system} in the zero-energy surface with $R^{*}$- or $R_1$-symmetry give rise, via the existence of EtoP connections, to BD-truncated homoclinic snaking. We now show that there are in fact infinitely many $R^{*}$-symmetric and $R_1$-symmetric periodic solutions with zero energy and, hence, many more families of homoclinic solutions of the equilibrium $\mathbf{0}$ with different symmetry properties. All of these homoclinic solutions indeed correspond to solitons of the GNLSE.

\begin{figure*}[hb!]
 {\includegraphics[width=\textwidth]{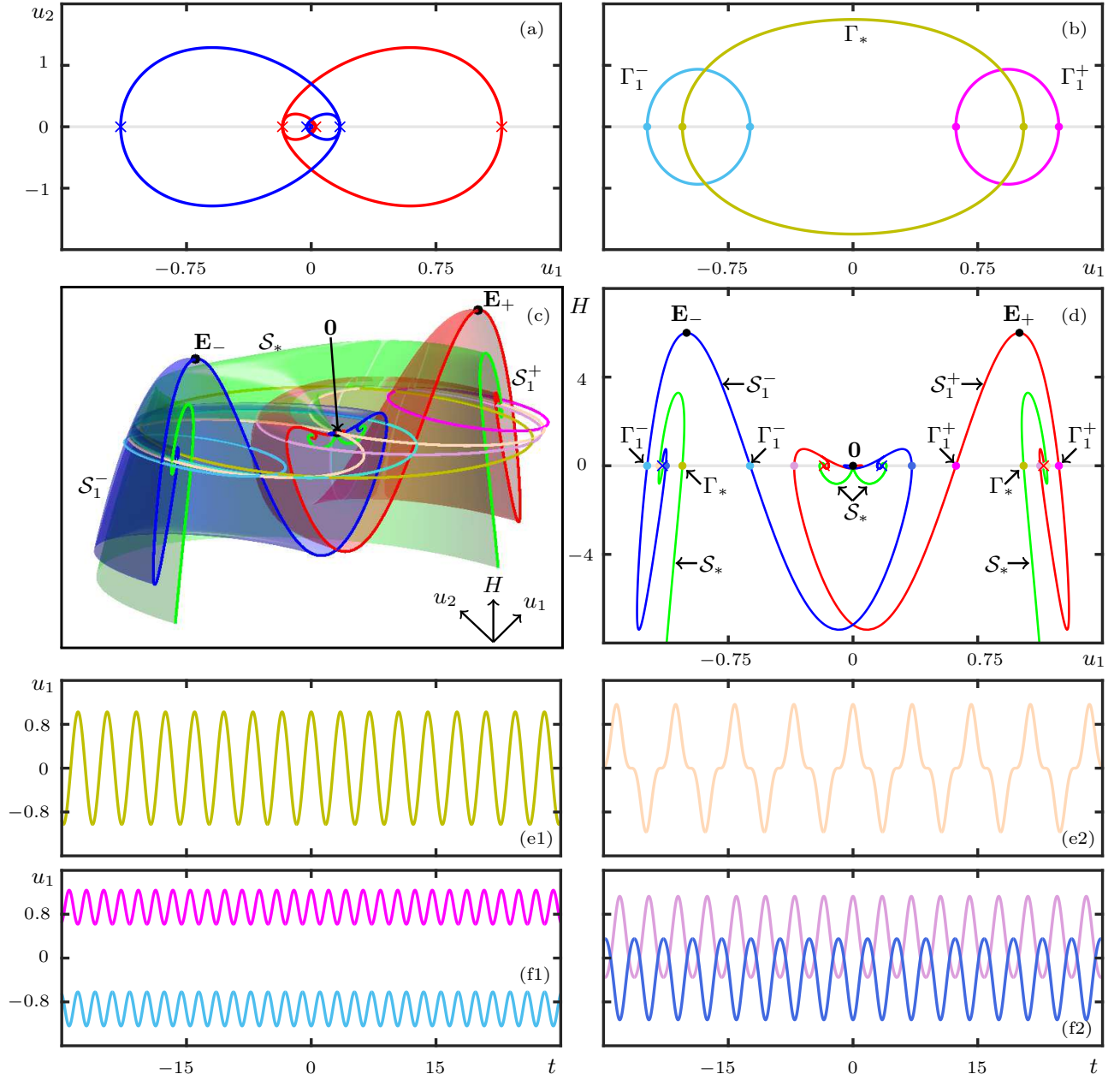}}
\caption{Existence of periodic solutions of system \eqref{system} for $(\beta_{2}, \beta_{4}, \gamma, \mu)=(0.4, -1, 1, 1)$. Panels (a) and (b) show, in the $(u_1,u_2)$-plane, the basic $R_1$-symmetric homoclinic orbit with its $R_2$-counterpart and the three periodic orbits $\Gamma_{*}$ and $\Gamma_{1}^{+/-}$ with $H(\mathbf{u})=0$. In both panels the light-grey line represents $u_2=0$, with crosses and circles indicating the intersections of the homoclinic and periodic orbits, respectively. Panel (c) shows, in $(u_{1},u_{2},H)$-space, half of the two-dimensional surfaces formed by three families of periodic solutions, referred to as $\mathcal{S}_1^{+}$ (red surface), $\mathcal{S}_1^{-}$ (blue surface) and $\mathcal{S}_*$  (green surface), with $R_1$- and $R^*$-symmetry, respectively. The black dots represent the three equilibria of system~\eqref{system}. Five different symmetric periodic orbits in the zero-energy level are also shown and their corresponding temporal profiles are plotted in panels (e1)-(f2). Panel (d) shows the intersection curves of the each surface with $u_2=0$; the light-grey line represents $H(\mathbf{u})=0$. The intersection points of each periodic solution in panels (c1)-(d2) are indicated by circles of the corresponding colour, and the intersection points of the basic homoclinic solution and its $R_1$-counterpart are indicated by red and blue crosses, respectively.} 
\label{fig:perorbits}
\end{figure*}

In what follows and specifically in Fig.~\ref{fig:perorbits}, we show representative solutions of system~\eqref{system} in phase space in different representations. Hence, periodic and homoclinic solutions now correspond to periodic and homoclinic orbits in phase space; see already panels~(a) and (b). These panels show in the $(u_1,u_2)$-plane, the basic $R_1$-symmetric homoclinic orbit and the periodic orbits we consider in Sec.~\ref{sec:connorbits}, for $(\beta_2,\beta_4,\mu,\gamma)=(0.4,-1,1,1)$. Notice that the basic $R_1$-symmetric homoclinic orbits come as a pair, and each of them spirals near the point $\mathbf{0}$ due to its complex conjugate eigenvalues. On the other hand, the periodic orbits in panel (b) are closed loops; the $R^*$-symmetric periodic orbit $\Gamma^{*}$ is a single loop, while the $R_1$-symmetry periodic orbits $\Gamma_1^{+/-}$ comes in pairs, which are each others $R_2$-counterparts.

Given that system \eqref{system} is reversible, for fixed values of the system parameters, periodic orbits come in one-parameter families; moreover, each homoclinic orbit gives rise to a one-parameter family of periodic orbits \cite{devaney1977blue, devaney1976}. Each periodic orbit can only be in a specific energy level $H$, which is why we consider them here as surfaces in $(u_{1},u_{2},H)$-space. This representation has the added advantage that it allows us to easily identify periodic orbits that are in the zero-energy level. Specifically, we consider three surfaces of periodic orbits, which we find by continuation in the energy $H$ from the primary homoclinic orbit  and its $R_2$-counterpart in Fig.~\ref{fig:perorbits}(a): individually these two homoclinic orbits give rise to $R_1$-symmetric periodic orbits each, which are each others $R_2$-counterparts, while their union gives rise to $R^*$-symmetric periodic orbits. 

Figure~\ref{fig:perorbits}(c) and (d) show these periodic orbits for $(\beta_2,\beta_4,\mu,\gamma)=(0.4,-1,1,1)$. Namely, panel~(c) shows three surfaces of periodic orbits in the $(u_{1},u_{2},H)$-space in a cut-away view that only shows their parts for positive $u_2$; note that the missing parts can be obtained by application of $R_1$, which is reflection in the $(u_{1},H)$-plane.  Here, in panel (c), the surfaces that contain the periodic orbits $\Gamma^{*}$ and $\Gamma_1^{+/-}$ are denoted $\mathcal{S}_{*}$ and $\mathcal{S}_{1}^{+/-}$, respectively. Also shown are six selected periodic orbits in the zero-energy level. Panel~(d) shows the respective intersection curves in the $(u_{1},H)$-plane, where the selected periodic orbits are identified as points with $H=0$. Panels~(e1) shows $\Gamma_{*}$ from Sec.~\ref{sec:RstarPeriodic} and panel~(f1) shows $\Gamma_{1}^{+/-}$ from Sec.~\ref{sec:R1Periodic}. Similarly, panels~(e2) and~(f2) show additional $R^*$-symmetric and $R_1$-symmetric periodic solutions, respectively.

Notice in Fig.~\ref{fig:perorbits}(c), and even more clearly in panel~(d), that the surface $\mathcal{S}_1^{+}$ has a global maximum in $H$ at $\bf{E_+}$, which lies in the $H(\mathbf{E}_+)$-energy level. Moreover, it has a global minimum in $H$ when it reaches a periodic orbit with $H(\mathbf{u}) \approx -7.4$. Thus, the Hamiltonian of this family of periodic orbits is bounded between these two values. The same statement is of course true for the surface $\mathcal{S}_1^{-}$, but its global maximum is the equilibrium $\mathbf{E}_-$, which is the $R_2$-counterpart of $\mathbf{E_+}$. The pair $\Gamma_1^{+/-}$ in panel~(f1) corresponds to the first intersection with $H(\mathbf{u})=0$ of this pair of surfaces when continued from $\mathbf{E_+}$ and $\mathbf{E}_-$. 

The surface $\mathcal{S}_{*}$ has a global maximum when it reaches a periodic orbit with $H(\mathbf{u})\approx 3.3$, but it does not have a global minimum in $H$; indeed, our numerical continuation results strongly suggest that this surface extends to any negative value of $H$. Note from Fig.~\ref{fig:perorbits}(c) and~(d) that, while its intersection with the $(u_{1},H)$-plane consists of three components, the surface $\mathcal{S}_{*}$ is nevertheless connected. The periodic solution $\Gamma_*$ in panel~(e1) is at the first intersection of this surface with $H(\mathbf{u})=0$, when continued for increasing $H$ from large negative values. 

The surfaces $\mathcal{S}^{+}$ and $\mathcal{S}^{-}$ accumulate on the basic $R_1$-symmetric homoclinic orbit and its $R_2$-counterpart, respectively. On the other hand, the surface $\mathcal{S}_{*}$  accumulate on the union of the basic $R_1$-symmetric homoclinic orbit and its $R_2$-counterpart. This is not so easy to see in the three-dimensional projection in Fig.~\ref{fig:perorbits}(c), but it can be observed more clearly in the $(u_1,H)$-plane in panel~(d). Notice that the respective intersection curves spiral into the intersection points of the two homoclinic orbits (marked by crosses), which means that these curves cross $H(\mathbf{u})=0$ infinitely often in the process. Hence, there are infinitely many additional periodic solutions with $R^{*}$- and $R_1$-symmetry in the zero-energy level. Panels~(e2) and~(f2) show the next such periodic solutions when continued on from the primary ones shown in panels~(e1) and~(f1), respectively. Each of these periodic orbits with $H(\mathbf{u})=0$ has a connection with $\mathbf{0}$ in certain parameter ranges. These infinitely many EtoP connections each give rise to BD-truncated homoclinic snaking scenarios with infinitley many homoclinic connections, with $R^{*}$- and $R_1$-symmetry of the kind we presented in Secs.~\ref{sec:RstarPeriodic} and~\ref{sec:R1Periodic}.

The picture that emerges from the discussion of only the surfaces $\mathcal{S}_*$ and $\mathcal{S}_1^{+/-}$ discussed here is indeed rather intriguing: each new homoclinic orbit gives rise to families of $R^{*}$- and $R_1$-symmetric periodic orbits, which, due to their spiralling create yet more periodic orbits in the zero-energy surface. Hence, there are infinite cascades of EtoP connections with infinitely many homoclinic orbits creating infinitely many new surfaces generating infinitely many periodic orbits each and so on. Moreover, there also exist more complicated connections associated with different periodic orbits, such as connections from a periodic orbit to itself (homoclinic orbits to a periodic orbit) and heteroclinic connections from one periodic orbit to another (PtoP connections) \cite{homburg2002multiple}. All these connections between periodic orbits also form more complex heteroclinic cycles. Therefore, there exist infinitely many additional homoclinic and periodic orbits that involve PtoP connections. How the different surfaces of periodic orbits are organised in phase space, and how this geometric structure changes as parameters are varied, is an interesting and challenging question. However, this is beyond the scope of this paper and will be discussed elsewhere.

\section{STABILITY PROPERTIES OF THE DIFFERENT TYPES OF SOLITONS}
\label{sec:simulations} 

The $R_1$-symmetric primary soliton from Fig.~\ref{fig:basic}(c) was considered by Tam \textit{et al}.~\cite{tam2018solitary,GDK,tam2019stationary} and found to be linearly stable in any parameter range. As we have just shown, there exist infinitely many other solitons with different symmetry properties over a broad parameter range of $\beta_2$, including for the case of a quartic fiber with $\beta_2=0$. It seems natural to suspect that all these other (multi-hump) solitons are linearly unstable. Determining the stability of soliton solution of a PDE is a challenging task \cite{sandstede2002stability}, and we restrict ourselves here to providing some first insights into the stability of the different types of solitons by means of simulations of the GNLSE with a split-step Fourier method (SSFM) \cite{agrawal2000nonlinear}. More specifically, we construct the respective soliton $u(t)$ from the particular homoclinic solution in Sec.~\ref{sec:connorbits}. We then perturb $u(t)$ in the same specific way as considered in ~\cite{tam2019stationary}, namely by increasing its size, here by $1\%$; that is, by considering the scaled profile $1.01\, u(t)$ as the input. We then evolve this perturbed profile with the SSFM with suitable accuracy settings to see how long it remains close to the initial constructed soliton, that is, propagates seemingly stably along the fibre before breaking up. In this way, we obtain an indication of whether and which multi-hump solitons might be observable in a physical experiment.

The simulations we performed are in no way exhaustive or representative of the different kinds of perturbations one may encounter in an experiment. Nevertheless, they do provide some insights into differences in stability of the various types of solitons. Perturbations of the $R_1$-symmetric primary soliton from Fig.~\ref{fig:basic}(c) die down during the simulation, meaning that this soliton can be propagated with the SSFM for an arbitrarily long distance along the fibre; this fact was used to determine suitable accuracy setting (determining time and space discretizations) for the SSFM. 

\begin{figure}[t!]
   \centering
   \includegraphics[scale=1.3]{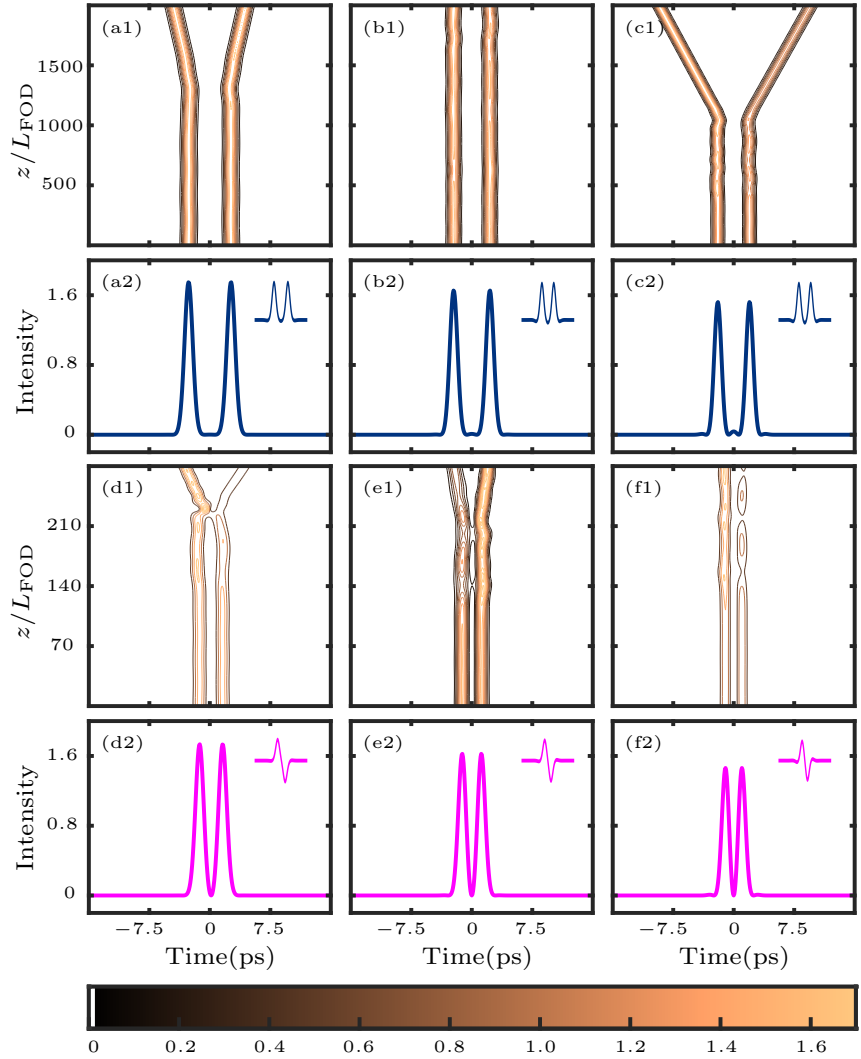}
    \caption{Evolution of the $1\%$-perturbed $R_1$-symmetric two-hump soliton in panels (a)--(c) and the $R_2$-symmetric primary soliton in panels (d)--(f), for fixed $(\beta_4,\mu,\gamma)=(-1,1,1)$ and $\beta_{2}=-0.2, \beta_{2}=0$ and $\beta_{2}=0.2$, respectively. For each case, the bottom panel shows the initial intensity profile, with the temporal trace of the corresponding homoclinic solutions in the top right, while the top panel shows the evolution as computed with the SSFM.}
\label{fig:simulations2}
\end{figure}

\begin{figure}[t!]
   \centering
   \includegraphics[scale=1.3]{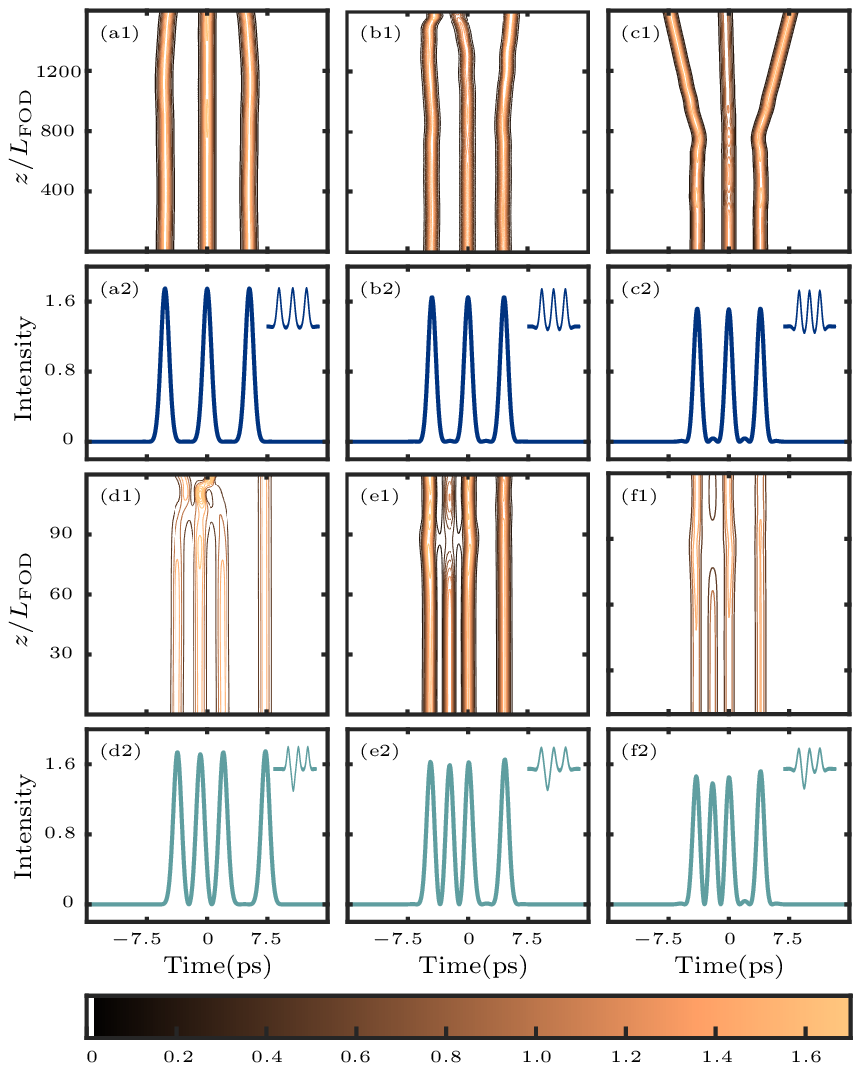}
    \caption{Evolution of the $1\%$-perturbed $R_1$-symmetric three-hump soliton in panels (a)--(c) and the corresponding symmetry-broken soliton in panels (d)--(f), for fixed $(\beta_4,\mu,\gamma)=(-1,1,1)$ and $\beta_{2}=-0.2, \beta_{2}=0$ and $\beta_{2}=0.2$, respectively. For each case, the bottom panel shows the initial intensity profile, with the temporal trace of the corresponding homoclinic solutions in the top right, while the top panel shows the evolution as computed with the SSFM.}
\label{fig:simulations3}
\end{figure}

Figures~\ref{fig:simulations2} and~\ref{fig:simulations3} show for fixed $(\beta_4,\mu,\gamma)=(-1,1,1)$ and for $\beta_{2}=-0.2, \beta_{2}=0$ and $\beta_{2}=0.2$ the intensity profiles of different types of $1\%$-perturbed solitons with their evolutions along the fibre as computed with the SSFM. More specifically, we consider the two-hump $R_1$-symmetric soliton in panels~(a)--(c) and the primary $R_2$-symmetric soliton in panels~(d)--(f) of Fig.~\ref{fig:simulations2}. Likewise, Fig.~\ref{fig:simulations3} shows the three-hump $R_1$-symmetric soliton in panels~(a)--(c) and the corresponding $R_1$-symmetry broken soliton in panels~(d)--(f). In each case, the bottom panel shows the initial intensity profile (at $z=0$), with the temporal trace of the corresponding homoclinic solution in the top-right corner; the top panel shows the evolution of the respective initial $1\%$-perturbed intensity profile.

As Figs.~\ref{fig:simulations2} and~\ref{fig:simulations3} show, all of these $1\%$-perturbed solitons break up after some finite distance, which supports the conjecture that they are indeed unstable. However, we find considerable differences between different initial solutions in how far along the fibre they can be propagated before breaking up. The differences manifest themselves for pure quartic solitons with $\beta_2=0$, the case that motivated our study, but also depend on the value $\beta_2$ of the quadratic dispersion. To quantify the effective distance over which a $1\%$-perturbed soliton may be observed in practice, we scale the fibre coordinate $z$ by the fourth-order dispersion length of a pulse given by
$$
L_{\rm{FOD}} = \frac{T^4_0}{|\beta_4|}.
$$
Here $T_0=\rm{FWHM}/ \, 2\sqrt{\log{2}}$ is determined from the full width at half maximum (FWHM) of the pulse; this is exact for a Gaussian (intensity) pulse, and a good approximation for pulses that are close to Gaussian \cite{agrawal2000nonlinear}. For initial $1\%$-perturbed multi-hump solitons in Figs.~\ref{fig:simulations2} and~\ref{fig:simulations3}, which all feature quite distictive pulses, we consider the largest pulse and find that a Gaussian is still a good fit; hence, we determine $L_{\rm{FOD}}$ from the FWHM of the largest pulse. We found that the values of the computed fourth-order dispersion length $L_{\rm{FOD}}$ of all initial solitons we considered agree up to two decimal places with that of the $R_1$-symmetric primary soliton, which we determined as $L_{\rm{FOD}} = 0.29$ for $\beta_2=-0.2$, $L_{\rm{FOD}} = 0.22$ for $\beta_2=0$ and $L_{\rm{FOD}} = 0.16$ for $\beta_2=0.2$. Furthermore, we computed from the FWHM the group velocity dispersion length $L_{\rm{GVD}} = T^2_0 / |\beta_2|$ (second order dispersion) for $\beta_2=-0.2$ and $\beta_2=0.2$. For all the initial $1\%$-perturbed solitons we consider $L_{\rm{GVD}}$ is one order of magnitude larger than $L_{\rm{FOD}}$. We conclude that the fourth-order dispersion is clearly dominant and, therefore, we use the above values of $L_{\rm{FOD}}$ to scale $z$ in Figs.~\ref{fig:simulations2} and \ref{fig:simulations3}.

The two-hump $R_1$-symmetric soliton in panels~(a)--(c) of Fig.~\ref{fig:simulations2} is only weakly unstable and can be observed over up to $1500\,L_{\rm{FOD}}$ in a quartic-dispersion fibre with $\beta_2=0$ before breaking up; with small second-order dispersion of $\beta_{2}=-0.2$ and $\beta_2=0.2$ this value drops to $1250\,L_{\rm{FOD}}$ and $1000\,L_{\rm{FOD}}$, respectively. By contrast, the primary $R_2$-symmetric soliton in panels~(d)--(f) of Fig.~\ref{fig:simulations2} is considerably more unstable and is observable only up to $120\,L_{\rm{FOD}}$ for the three considered values of $\beta_2$; beyond this value of $L_{\rm{FOD}}$ we observe noticeable interactions between the two intensity pulses and pulse break-up shorty thereafter. As panels~(a)--(c) of Fig.~\ref{fig:simulations3} show, the perturbed three-hump $R_1$-symmetric soliton is observable considerably farther along the fibre, up to about $600\,L_{\rm{FOD}}$ when pulse interactions become visible for all $\beta_2$; note that the negative $\beta_2$ increases the observation length, while positive $\beta_2$ decreases it. The related $R_1$-symmetry broken soliton in panels~(d)--(f) is clearly much more unstable and can be observed only up to about $60\,L_{\rm{FOD}}$. 

Our preliminary simulation results suggest that $R_1$-symmetric solitons with up to three humps appear to be only weakly unstable. This means that, when launched quite precisely into an actual quartic dispersion fibre, they might be sustained over a sufficiently large number of dispersion lengths to be observable. Confirming this in an experiment is clearly a considerable challenge. Solutions without $R_1$-symmetry, on the other hand, appear to be much more unstable, and are unlikely to be observable experimentally. Note that Figs.~\ref{fig:simulations2} and~\ref{fig:simulations3} suggest that the different initial profiles may break up differently, by their pulses interacting in different ways. The further study of such instabilities via the repulsion and/or attraction of neighbouring pulse is a challenging subject for future research.

\section{CONCLUSIONS}
\label{sec:conclusions}

We investigated the existence of solitons of the GNLSE in the presence of both quartic and quadratic dispersion. Taking a dynamical system approach, we made a traveling wave ansatz to translate solitons of the GNLSE into homoclinic solutions to the equilibrium $\mathbf{0}$ of system~\eqref{system}, which is Hamiltonian and features two reversible symmetries, $R_1$ and $R_2$. We found that for both signs of the quadratic dispersion $\beta_2$ there exist infinitely many homoclinic solutions of system~\eqref{system} with different symmetry properties, which correspond to infinitely many solitons of the GNLSE. Each familly of homoclinic solutions is associated with a heteroclinic cycle formed by different EtoP connections between $\mathbf{0}$ and a specific periodic solution in the zero-energy level. The symmetries of the periodic solution determine what families of homoclinic solutions they generate.

We presented here four different families of homoclinic solutions, namely those with $R_1$- and $R_2$-symmetry, as well as related $R_1$-symmetry broken and $R_2$-symmetry broken homoclinic solutions.  Both the $R_1$-symmetric and $R_2$-symmetric primary homoclinic solutions emerge from a Hamiltonian-Hopf bifurcation and exists over the entire $\beta_2$-interval where $\mathbf{0}$ has complex conjugates eigenvalues with non-zero real part. All the other $R_1$- and $R_2$-symmetric and associated symmetry broken multi-hump homoclinic solutions, on the other hand, come as pairs on branches that meet at fold points for particular values of $\beta_2$; the respective fold points accumulate on the $\beta_2$-values of folds of the corresponding heteroclinic EtoP connections between $\mathbf{0}$ and periodic solutions in the zero-energy level.

These results were obtained by combining the theory of four-dimensional reversible Hamiltonian system with state-of-the-art continuation techniques that enabled us to compute branches of homoclinic solutions, as well as those of the corresponding EtoP connections that organise them. In this way, we provided numerical evidence for the overall organisation of homoclinic solutions, which all emerge/disappear in a Belyakov-Devaney (BD) bifurcation, leading to a structure, also reported in the LL-equation \cite{articleParra}, that we refer to as BD-truncated homoclinic snaking. We also showed that there exist infinitely many periodic solutions that generate EtoP connections and, hence, associated families of multi-hump homoclinic solutions; moreover, we presented the regions of existence of all these different solitons in the $(\beta_2, \beta_4)$- and the $(\beta_2, \mu)$-plane. Overall, our results provide guidance in the form of a \enquote{road map} of the plethora of solitons exhibited by the GNLSE, and how they are organised in families by EtoP cycles. 

Finally, in a first consideration of the stability properties of the various multi-hump solitons, we checked how far along the fibre a specific perturbation of such a soliton propagates before breaking up. Our simulation results indicate that all solitons, except the $R_1$-symmetric single pulse soliton, are unstable; moreover the $R_1$-symmetric double-hump soliton and the $R_1$-symmetric triple-hump soliton appear to be only weakly unstable: (after perturbation) they can be progagated by numerical simulation of the GNLSE over a considerable number of dispersion lengths. Hence, when launched carefully into the right kind of wave guide, they might be observable experimentally. Sustaining such weakly unstable solitons over sufficiently large distances in an experiment is indeed a considerable challenge.

A number of interesting theoretical questions arise from our study. First of all, we have observed that, as $\beta_2 $ is varied, different surfaces of periodic orbits may interact and bifurcate, which leads to changes of the types of periodic orbits that can be found in the zero-energy level. In turn, this influences the structure of available EtoP connections that organise families of homoclinic orbits. Moreover, we have evidence that connections between the same and/or between different periodic orbits, referred to as PtoP cycles, exist in the zero-energy surface. Hence, more complicated heteroclinic cycles from $\mathbf{0}$  back to itself can be constructed, which presumably generate associated families of homoclinic solutions. The study of these additional connections and associated types of solitons of the GNLSE is the subject of ongoing work. The stability analysis of the different solitons as solutions of the GNLSE, that is, of the PDE, remains a considerable challenge. Finally, recent experiments  \cite{PhysRevResearch.3.013166} have shown the feasibilities of creating waveguides with higher even order dispersions, such as sextic $(\beta_6)$, octic $(\beta_8)$, and decic $(\beta_{10})$ dispersions. Our analysis constitutes the first step towards understanding the existence of solitons for different configuration of such higher-order dispersion terms.

\section*{ACKNOWLEDMENTS}

We would like to thank C. Martijn De Sterke and Kevin Tam for several helpful discussions.


\providecommand{\noopsort}[1]{}\providecommand{\singleletter}[1]{#1}%

\end{document}